\documentclass[preprint]{revtex4}
\usepackage{graphicx}
\usepackage{psfrag}
\usepackage{color}
\usepackage{ulem}
\footnotesize
\usepackage{booktabs}
\usepackage{hyperref}
\usepackage{mathtools}
\usepackage{natbib}
\begin{document}
	
\title{Relations between generalized parton distributions  and transverse momentum dependent parton distributions }

\author{Bheemsehan Gurjar$^1$, Dipankar Chakrabarti$^1$, Poonam  Choudhary$^1$, Asmita Mukherjee$^2$, Pulak Talukdar$^1$}

\affiliation{$^1$Indian Institute of Technology Kanpur, Kanpur-208016, India}	
\affiliation{$^2$Indian Institute of Technology Bombay, Powai, Mumbai 400076, India}

\begin{abstract}
We investigate the relations between transverse momentum dependent parton distributions (TMDs) and generalized parton distributions (GPDs)  in a light-front quark-diquark model motivated by soft wall AdS/QCD. Many relations are found to have similar structure in different models.  It is found that a relation between the Sivers function and the GPD $E_q$ can be obtained in this model in terms of a lensing function.  The quark orbital angular momentum is calculated and the results are compared with the results in other similar models.  Implications of the results are discussed. Relations among different TMDs in the model are also presented.

\end{abstract}
\maketitle

\section{Introduction}
Understanding the structure of the nucleon  in terms of its fundamental constituents, quarks and gluons, in three dimension has attracted quite a lot of interest in hadron physics  in recent days.  These are  investigated in terms of the different distributions of quarks and gluons that encode their internal dynamics, as well as the correlations between the intrinsic momentum and spin. 
The parton distributions are  probed traditionally  in high energy scattering experiments, where the interactions take place through the quarks and gluons, and the scattering cross section depends on the probability to find a quark with momentum fraction $x$ inside the parent nucleon at a given momentum scale (energy of the experiment).  These are called collinear parton distributions as they are not sensitive to the intrinsic transverse momentum of the quarks and gluons.  However, single spin asymmetries observed in semi-inclusive deep inelastic scattering (SIDIS) or Drell-Yan (DY) processes, where the target or one of the proton beams is polarized, depend on transverse momentum dependent parton distribution (TMDs) \cite{Collins:2011zzd}; that give the distribution of quarks and gluons in three dimensional momentum space. The TMDs are functions of the longitudinal momentum fraction $x$ and transverse momentum $k_T$ of the partons. There are eight leading quark TMDs for the proton, each encode a different momentum-momentum or momentum-spin correlation. These TMDs can be expressed in terms  of the quark field operators, and in order to have color gauge invariance one needs the inclusion of a Wilson line or gauge link. Also known as the initial and/or final  state interaction, these basically resum the soft gluon exchanges between the hard part and the soft part of the process.  The gauge links are process dependent and thus they introduce a process dependence in the TMDs. For the so-called time reversal odd (T-odd) TMDs like Sivers function  or Boer-Mulders function  the inclusion of the gauge link is essential \cite{Boer:2003cm}.

Another set of observables that have gained a lot of interest in recent days are  the   generalized parton distributions (GPDs) of the nucleon \cite{Diehl:2003ny}. These are  probed in exclusive processes like the deeply virtual Compton scattering (DVCS)  or the  deeply virtual meson production. GPDs can be expressed in terms of off-forward matrix element of a bilocal operator, and these do not have probabilistic interpretation. There are eight leading GPDs for the quarks. In the forward limit or when the momentum transfer in the process is zero, GPDs reduce to the collinear pdfs; whereas integrating them over $x$ one gets the form factors. When the momentum transfer $\Delta_T$ is purely in the transverse direction, by taking a Fourier transform with respect to $\Delta_T$ one obtains impact parameter dependent parton distributions (IPDpdfs) \cite{Burkardt:2002hr}  that are functions of $x$ and transverse impact parameter $b_T$. These have a probabilistic interpretation : they give the distribution of quarks with longitudinal momentum fraction $x$ in $b_T$ plane. 

As a matter of fact, here is no direct one-to-one correspondence between the TMDs and GPDs. This is because $b_T$ and $k_T$ are not Fourier conjugate variable to each other. $b_T$  is the Fourier conjugate to the momentum transfer $\Delta_T$, and $k_T$ may be interpreted as the average momentum of the active quark. However, $b_T$ and $k_T$ obey Heisenberg's uncertainty principle as the corresponding operators do not commute. So although both the GPDs and TMDs give  three dimensional quark-gluon picture of the nucleon, one is not related to the other through a Fourier transform. A model independent connection between the GPDs and the TMDs  can be obtained through the generalized transverse momentum dependent pdfs (GTMDs)   or Wigner functions, which are Fourier transforms of the GTMDs \cite{Lorce:2011kd, Meissner:2009ww}. These give the most general tomographic picture of the nucleon. Integration over $b_T$ sets $\Delta_T=0$ and the Wigner distributions become the TMD correlators. On the other hand, integrating over $k_T$ sets $z_T=0$ and they reduce to IPDPdfs.  

However, in some models, certain relations between the TMDs and GPDs are found to hold. One such relation is between the Sivers function and the GPD $E_q$ in impact parameter space, in some particular model where the final state interaction can be factored out in what is called 'chromodynamic lensing function' \cite{Burkardt:2003uw, Burkardt:2003yg}.  This gives an intuitive picture of Sivers effect in such models in terms of distortions in transverse impact parameter space, due to the non-zero  orbital angular momentum of the active quark. Such relations do not hold if higher order corrections are included.  Thus, model dependent relations are important as they help to understand  the physics related to these TMDs in the framework of the effective theory on which such models are based.  In \cite{Meissner:2007rx} a systematic study of all possible non-trivial model-dependent relations between TMDs and GPDs was performed and such relations were arranged in four categories depending on the number of derivatives in impact parameter space.  In \cite{Pasquini:2019evu}  it was shown  that the relation connecting a T-odd TMD to a distortion in impact parameter space through a lensing function holds only in models where the nucleon is described  as a two-particle bound state, like a quark and a diquark. Further, such relations are not satisfied if axial vector diquarks are included in the model. 

In the present work, we investigate model relations in a light-front diquark model where the analytic form of the light-front wave functions (LFWF) is motivated by the AdS/QCD correspondence. The model includes both scalar and axial vector diquarks \cite{PhysRevD.94.094020}. Total  nucleon wavefunctions  are obtained by  the LF holographic wavefunction multiplied by the momentum dependent helicity wave functions. Finally, we incorporate the final state interaction in the wavefunctions to evaluate the T-odd TMDs.   We also calculate the orbital angular momentum of the quarks in this model and discuss the results.

 The plan of the paper is as follows.  In Sec,\ref{model}, we introduce the light front  quark-diquark model used in this work. The modification of the wave functions to incorporate the FSI effect is given in Sec.\ref{model_fsi}.Then  in Sec.\ref{defs}, we define the GPDs and TMDs in this model.  Model independent relations among the GPDs and TMDs are also presented in this section. Then we present  the model dependent relations between the GPDs and TMDs in Sec.\ref{relations1}. The model result of the lensing function is discussed in Sec.\ref{lensing} and the relations among different TMDs are presented in Sec.\ref{relations2}.  In Sec.\ref{oam}, different definitions of orbital angular momentum in terms of GPDs and TMDs are evaluated. Finally, we conclude the paper in Sec.\ref{concl}.

	\section{Light Front Quark Diquark Model}\label{model}
	
Here we briefly introduce the  light-front quark-diquark model developed in \cite{PhysRevD.89.054033, PhysRevD.94.094020} with the wave functions modelled from the effective two-particle wave function predicted by AdS/QCD. The particular model  of nucleons employed in this work is  considered to be a linear combination of a quark-diquark state including both the scalar diquark and axial-vector diquarks \cite{PhysRevD.94.094020}. With the $SU(4)$ spin-flavor structure, the proton states can be written as \cite{Bacchetta:2008af}
	\begin{align} \label{eq:1}
	|P ; \pm\rangle=C_{\mathrm{S}}\left|u S^{0}\right\rangle^{\pm}+C_{\mathrm{V}}\left|u A^{0}\right\rangle^{\pm}+C_{\mathrm{VV}}\left|d A^{1}\right\rangle^{\pm}.
	\end{align}
S and A represent the scalar and axial-vector diquarks with isospin at their superscripts. Under the isospin symmetry, the neutron state can be obtained from the above expression (Eq.(\ref{eq:1})) with the interchange of  $u\leftrightarrow d$.
The two particle Fock-state expansion for $J^{z}=\pm\frac{1}{2}$ for spin-0	diquark is given by 
\begin{align} \label{eq:2}
|u S\rangle^{\pm}= \int \frac{\mathrm{d} x \mathrm{~d}^{2} \mathbf{p}_{T}}{2(2 \pi)^{3} \sqrt{x(1-x)}}\left[\psi_{+}^{\pm(u)}\left(x, \mathbf{p}_{T}\right)\left|+\frac{1}{2}, 0 ; x P^{+}, \mathbf{p}_{T}\right\rangle\right. \nonumber \\
\left.+\psi_{-}^{\pm(u)}\left(x, \mathbf{p}_{T}\right)\left|-\frac{1}{2}, 0 ; x P^{+}, \mathbf{p}_{T}\right\rangle\right],
\end{align}
where  $|\lambda_{q},0;xP^{+},\mathbf{k}_{T}\rangle$  represents the two particle Fock state with active quark  with helicity $\lambda_{q}=\pm \frac{1}{2}$  and carrying longitudinal momentum $xP^+$ and transverse momentum $\mathbf{k}_T$   and a scalar diquark with helicity $\lambda_{S}=0$. Similarly the state with spin-1 diquark is given as 
\begin{equation}\label{eq:3}
\begin{aligned}
|v A\rangle^{\pm}=& \int \frac{\mathrm{d} x \mathrm{~d}^{2} \mathbf{p}_{T}}{2(2 \pi)^{3} \sqrt{x(1-x)}}\left[\psi_{++}^{\pm(v)}\left(x, \mathbf{p}_{T}\right)\left|+\frac{1}{2}+1 ; x P^{+}, \mathbf{p}_{T}\right\rangle\right.\\
&+\psi_{-+}^{\pm(v)}\left(x, \mathbf{p}_{T}\right)\left|-\frac{1}{2}+1 ; x P^{+}, \mathbf{p}_{T}\right\rangle 
+\psi_{+0}^{\pm(v)}\left(x, \mathbf{p}_{T}\right)\left|+\frac{1}{2} 0 ; x P^{+}, \mathbf{p}_{T}\right\rangle \\
&+\psi_{-0}^{\pm(v)}\left(x, \mathbf{p}_{T}\right)\left|-\frac{1}{2} 0 ; x P^{+}, \mathbf{p}_{T}\right\rangle 
+\psi_{+-}^{\pm(v)}\left(x, \mathbf{p}_{T}\right)\left|+\frac{1}{2}-1 ; x P^{+}, \mathbf{p}_{T}\right\rangle \\
&\left.+\psi_{--}^{\pm(v)}\left(x, \mathbf{p}_{T}\right)\left|-\frac{1}{2}-1 ; x P^{+}, \mathbf{p}_{T}\right\rangle\right],
\end{aligned}
\end{equation}
where $|\lambda_{q},\lambda_{A}; xP^{+}, \mathbf{p}_{T}\rangle$ represents a two-particle state with a quark of helicity $\lambda_{q}=\pm\frac{1}{2}$ and an axial-vector diquark with helicity $\lambda_{A}=\pm 1,0$(triplet).

\subsection{Final state interaction and T-odd TMDs}\label{model_fsi}
The final state interactions (FSI) \cite{Brodsky_2002} provide a non-trivial phase in the amplitude which is required to produce non-vanishing spin asymmetries observed in SIDIS processes.     In this model, the contribution of the FSI  is included in the light-front wave functions \cite{Hwang:2010dd} to evaluate the leading twist T-odd TMDs, i.e., the Sivers function, $f_{1 T}^{\perp q}(x,\mathbf{p}_{T}^{2})$  and the Boer-Mulders function, $h_{1}^{\perp q}(x,\mathbf{p}_{T}^{2})$.   The modified wave functions\cite{Maji:2017wwd}  are given by 
\\
(a) for scalar diquark
\begin{align}\label{eq:4}
\psi_{+}^{+(u)}\left(x, \mathbf{p}_{T}\right)={}&N_{\mathrm{S}}\left[1+i \frac{e_{1} e_{2}}{8 \pi}\left(\mathbf{p}_{T}^{2}+B\right) g_{1}\right] \varphi_{1}^{(u)}\left(x, \mathbf{p}_{T}\right) \nonumber \\
\psi_{-}^{+(u)}\left(x, \mathbf{p}_{T}\right)={}& N_{\mathrm{S}}\left(-\frac{p^{1}+i p^{2}}{x M}\right)
\left[1+i \frac{e_{1} e_{2}}{8 \pi}\left(\mathbf{p}_{T}^{2}+B\right) g_{2}\right] \varphi_{2}^{(u)}\left(x, \mathbf{p}_{T}\right) \nonumber \\
\psi_{+}^{-(u)}\left(x, \mathbf{p}_{T}\right)={}& N_{\mathrm{S}}\left(\frac{p^{1}-i p^{2}}{x M}\right)
\left[1+i \frac{e_{1} e_{2}}{8 \pi}\left(\mathbf{p}_{T}^{2}+B\right) g_{2}\right] \varphi_{2}^{(u)}\left(x, \mathbf{p}_{T}\right) \nonumber \\
\psi_{-}^{-(u)}\left(x, \mathbf{p}_{T}\right)={}& N_{\mathrm{S}}\left[1+i \frac{e_{1} e_{2}}{8 \pi}\left(\mathbf{p}_{T}^{2}+B\right) g_{1}\right] \varphi_{1}^{(u)}\left(x, \mathbf{p}_{T}\right)
\end{align}
(b) for axial-vector diquark 
(for J = +1/2)
\begin{align} \label{eq:5}
{}&\psi_{++}^{+(\nu)}\left(x, \mathbf{p}_{T}\right)=N_{1}^{(\nu)} \sqrt{\frac{2}{3}}\left(\frac{p^{1}-i p^{2}}{x M}\right)\left[1+i \frac{e_{1} e_{2}}{8 \pi}\left(\mathbf{p}_{T}^{2}+B\right) g_{2}\right] \varphi_{2}^{(\nu)}\left(x, \mathbf{p}_{T}\right) \nonumber \\
{}&\psi_{-+}^{+(\nu)}\left(x, \mathbf{p}_{T}\right)=N_{1}^{(\nu)} \sqrt{\frac{2}{3}}\left[1+i \frac{e_{1} e_{2}}{8 \pi}\left(\mathbf{p}_{T}^{2}+B\right) g_{1}\right] \varphi_{1}^{(\nu)}\left(x, \mathbf{p}_{T}\right) \nonumber \\
{}&\psi_{+0}^{+(\nu)}\left(x, \mathbf{p}_{T}\right)=-N_{0}^{(\nu)} \sqrt{\frac{1}{3}}\left[1+i \frac{e_{1} e_{2}}{8 \pi}\left(\mathbf{p}_{T}^{2}+B\right) g_{1}\right] \varphi_{1}^{(\nu)}\left(x, \mathbf{p}_{T}\right) \nonumber \\
{}&\psi_{-0}^{+(\nu)}\left(x, \mathbf{p}_{T}\right)=N_{0}^{(\nu)} \sqrt{\frac{1}{3}}\left(\frac{p^{1}+i p^{2}}{x M}\right)\left[1+i \frac{e_{1} e_{2}}{8 \pi}\left(\mathbf{p}_{T}^{2}+B\right) g_{2}\right] \varphi_{2}^{(\nu)}\left(x, \mathbf{p}_{T}\right) \nonumber \\
{}&\psi_{+-}^{+(\nu)}\left(x, \mathbf{p}_{T}\right)=0, \nonumber \\
{}&\psi_{--}^{+(\nu)}\left(x, \mathbf{p}_{T}\right)=0
\end{align}
and, (for J = -1/2)
\begin{align}  \label{eq:6}
{}&\psi_{++}^{-(\nu)}\left(x, \mathbf{p}_{T}\right)=0,\nonumber \\
{}&\psi_{-+}^{-(\nu)}\left(x, \mathbf{p}_{T}\right)=0 \nonumber \\
{}&\psi_{+0}^{-(\nu)}\left(x, \mathbf{p}_{T}\right)=N_{0}^{(\nu)} \sqrt{\frac{1}{3}}\left(\frac{p^{1}-i p^{2}}{x M}\right)\left[1+i \frac{e_{1} e_{2}}{8 \pi}\left(\mathbf{p}_{T}^{2}+B\right) g_{2}\right] \varphi_{2}^{(\nu)}\left(x, \mathbf{p}_{T}\right) \nonumber \\
{}&\psi_{-0}^{-(\nu)}\left(x, \mathbf{p}_{T}\right)=N_{0}^{(\nu)} \sqrt{\frac{1}{3}}\left[1+i \frac{e_{1} e_{2}}{8 \pi}\left(\mathbf{p}_{T}^{2}+B\right) g_{1}\right] \varphi_{1}^{(\nu)}\left(x, \mathbf{p}_{T}\right) \nonumber \\
{}&\psi_{+-}^{-(\nu)}\left(x, \mathbf{p}_{T}\right)=-N_{1}^{(\nu)} \sqrt{\frac{2}{3}}\left[1+i \frac{e_{1} e_{2}}{8 \pi}\left(\mathbf{p}_{T}^{2}+B\right) g_{1}\right] \varphi_{1}^{(\nu)}\left(x, \mathbf{p}_{T}\right) \nonumber \\
{}&\psi_{--}^{-(\nu)}\left(x, \mathbf{p}_{T}\right)=N_{1}^{(\nu)} \sqrt{\frac{2}{3}}\left(\frac{p^{1}+i p^{2}}{x M}\right)\left[1+i \frac{e_{1} e_{2}}{8 \pi}\left(\mathbf{p}_{T}^{2}+B\right) g_{2}\right] \varphi_{2}^{(\nu)}\left(x, \mathbf{p}_{T}\right)
\end{align}
Where, 
\begin{eqnarray}
g_{1} =\int_{0}^{1} d \alpha \frac{-1}{\alpha(1-\alpha) \mathbf{p}_{T}^{2}+\alpha m_{g}^{2}+(1-\alpha) B} \\
g_{2} =\int_{0}^{1} d \alpha \frac{-\alpha}{\alpha(1-\alpha) \mathbf{p}_{T}^{2}+\alpha m_{g}^{2}+(1-\alpha) B}
\end{eqnarray}
and, 
\begin{eqnarray}
B=x(1-x)(-M^{2}+\frac{m_{q}^{2}}{x}+\frac{m_{D}^{2}}{1-x}). \label{Bx}
\end{eqnarray}
Here M, $m_{q}$, $m_{D}$ and $m_{g}$ are the masses of the proton, struck quark, diquark and the gluon respectively. $e_{1}$ and $e_{2}$ are the color charges of the struck quark and diquark, and FSI gauge exchange strength is $\frac{e_{1}e_{2}}{4\pi}$. We take $m_{g}=0$ at the end of the calculations. $N_{s}, N_{0}^{\nu}$ and $N_{1}^{\nu}$ are the normalisation constants. The LFWFs $\varphi_{i}^{\nu}(x,\mathbf{p}_{T})$ are modified from the soft-wall AdS/QCD predictions as \cite{ PhysRevD.94.094020}
\begin{align}\label{eq:8}
\varphi_{i}^{(\nu)}\left(x, \mathbf{p}_{T}\right)=A_{i}^{\nu}(x)\exp[-a(x)\mathbf{p}_{T}^{2}]
\end{align}
With 
\begin{eqnarray}
	A_{i}^{\nu}(x)=\frac{4\pi }{\kappa }\sqrt{\frac{\log\left(\frac{1}{x}\right)}{1-x}}x^{a_{i}^{\nu}}(1-x)^{b_{i}^{\nu}},
\end{eqnarray}
and 
\begin{eqnarray}
	a(x)= \delta^\nu \frac{\log(\frac{1}{x})}{\kappa ^2(1-x)^2}.  \label{ax}
\end{eqnarray}
The  parameters  $a_{i}^{\nu}$, $ b_{i}^{\nu}$ and  $\delta^{\nu}$ are obtained by  fitting the electromagnetic form factors.   The wavefunction (Eq.(\ref{eq:8}))  reduces to the AdS/QCD prediction \cite{Brodsky:2007hb} for the parameters $a_{i}^{\nu}=b_{i}^{\nu}=0$ and $\delta^{\nu}=1$. 
We use the AdS/QCD scale parameter $\kappa=0.4 GeV$ \cite{Chakrabarti:2013gra} and the quark masses are assumed to be zero.  For completeness and to be self-contained, we list the parameters obtained in Ref.\cite{ PhysRevD.94.094020}. The normalization constants are $N_{s}=2.0191$, $N_{0}^{u}=3.2050$, $N_{0}^{d}=5.9423$, $N_{1}^{u}=0.9895$, $N_{1}^{d}=1.1616$, $C_{s}^{2}=1.3872$, $C_{V}^{2}=0.6128$ and $C_{VV}^{2}=1.0$ and the other parameters are listed in Table \ref{table1}. Henceforth, we refer this model as LFQDQ model.
\begin{table} [ht]
\centering
\begin{tabular}{ c c c c c c c}
\hline \hline
 $\nu$  &$a_{1}^{\nu}$ & $b_{1}^{\nu}$ & $a_{2}^{\nu}$ & $b_{2}^{\nu}$ & $\delta^{\nu}$ \\ [0.5ex]
	\hline $u$ &  $0.280 \pm 0.001$ & $0.1716 \pm 0.0051$ & $0.84 \pm 0.02$ & $0.2284 \pm 0.0035$ & 1.0 \\
	$d$ & $0.5850 \pm 0.0003$ & $0.7000 \pm 0.0002$ & $0.9434_{-0.0013}^{+0.0017}$ & $0.64_{-0.0022}^{+0.0082}$ & 1.0 \\
\hline \hline
\end{tabular}
\caption{In the light-front AdS/QCD axial-vector diquark model, the values of the fitted parameters for u and d quarks at $\mu_{0}=0.313$ GeV.}
	\label{table1}
\end{table}

	\section{Definitions and Trivial Relations} \label{defs}

	\subsection{Generalized parton distributions}  
	
	The GPDs(for a review, see Ref. \cite{Diehl:2003ny})
	are defined as off-forward matrix elements of the bilocal operator of light-front correlation functions of vector, axial
	vector, and tensor current. The off-forward correlator is given by 
	\begin{eqnarray}
		F^{q[\Gamma]}\left(x, \Delta ; \lambda, \lambda^{\prime}\right) 
		&=&\frac{1}{2} \int \frac{d z^{-}}{2 \pi} e^{i k \cdot z}\left\langle p^{\prime} ; \lambda^{\prime}\right| \bar{\psi}\left(-\frac{1}{2} z\right) \Gamma
		\quad \nonumber\\
		&& \times\left.\mathcal{W}_{GPD}\left(-\frac{1}{2} z ; \frac{1}{2} z\right) \psi\left(\frac{1}{2} z\right)|p ; \lambda\rangle\right|_{z^{+}=0^{+},\mathbf{z}_{T}=\mathbf{0}_{T}},  \label{gpdcorrelator}
	\end{eqnarray}
	where $p(p^{\prime})$ and $\lambda(\lambda^{\prime})$ denote the momenta and the
	helicity of the initial (final) state of proton, respectively. The object $\Gamma$
	is a generic matrix in Dirac space, at leading twist, it can be $\gamma^+, \gamma^+\gamma^5$ or $\sigma^{+\perp} \gamma^5$ . The Wilson line $\mathcal{W}_{GPD}$   is required for  gauge invariance of the
	correlator   and is given by
	\begin{eqnarray} \label{gpdwilson}
		\left.\mathcal{W}_{GPD}\left(-\frac{1}{2} z ; \frac{1}{2} z\right)\right|_{z^{+}=0^{+},\mathbf{z}_{T}=\mathbf{0}_{T}}
		=\left[0^{+},-\frac{1}{2} z^{-}, \mathbf{0}_{T} ; 0^{+}, \frac{1}{2} z^{-}, \mathbf{0}_{T}\right] \nonumber \\
		=\mathcal{P} \exp \left(-i g \int_{-(1 / 2) z^{-}}^{(1 / 2) z^{-}} d y^{-} t_{a} A_{a}^{+}\left(0^{+}, y^{-}, \mathbf{0}_{T}\right)\right)
	\end{eqnarray}
	where $\mathcal{P}$ is the path ordering and $t_{a}$ represents the Gell-Mann matrices. For the three particular $\Gamma$ in Eq. (\ref{gpdcorrelator}) we can obtain the leading twist GPDs \cite{Meissner:2007rx} as
	\begin{eqnarray}
		F^{q[\gamma^{+}]}\left(x, \Delta ; \lambda, \lambda^{\prime}\right) 
		&=& \frac{1}{2 P^{+}} \bar{u}\left(p^{\prime}, \lambda^{\prime}\right)\left[H^{q} \gamma^{+}+E^{q} \frac{i}{2 M} \sigma^{+\alpha} \Delta_{\alpha}\right] u(p, \lambda), \label{gpd1} \\
		F^{q[\gamma^{+} \gamma_{5}]}\left(x, \Delta ; \lambda, \lambda^{\prime}\right) 
		&=& \frac{1}{2 P^{+}} \bar{u}\left(p^{\prime}, \lambda^{\prime}\right)\left[\tilde{H}^{q} \gamma^{+} \gamma_{5}+\tilde{E}^{q} \frac{\gamma_{5} \Delta^{+}}{2 M}\right] u(p, \lambda), \label{gpd2}\\
		F^{q[\sigma^{+j} \gamma_{5}]}\left(x, \Delta ; \lambda, \lambda^{\prime}\right) 
		&=& \frac{1}{2 P^{+}} \bar{u}\left(p^{\prime}, \lambda^{\prime}\right) [ H_{T}^{q} \sigma^{+j} \gamma_{5}+\tilde{H}_{T}^{q} \frac{\epsilon^{+j \alpha \beta} \Delta_{\alpha} P_{\beta}}{M^{2}}+E_{T}^{q} \frac{\epsilon^{+j\alpha\beta}\Delta_{\alpha}\gamma_{\beta}}{2M} \nonumber \\
		&& + \tilde{E}_{T}^{q} \frac{\epsilon^{+j\alpha\beta}P_{\alpha}\gamma_{\beta}}{M} ] u(p,\lambda).	\label{gpd3}
			\end{eqnarray}
	Where M denotes the mass of the proton and $j= 1, 2$ is a transverse
	index, $P=(p+p^{\prime})/2$ denotes the average nucleon momentum and $\Delta=p-p^{\prime}$ is the momentum transfer to the nucleon. The $H$ and $E$, so called unpolarized GPDs and the
	helicity dependent GPDs, $\tilde{H}$ and $\tilde{E}$ are chiral-even, while
	$H_{T}$, $\tilde{H}_{T}$, $E_{T}$, and $\tilde{E}_{T}$ are chiral-odd. There exist eight leading twist quark GPDs. All GPDs are real valued which follows from  time-reversal and depend on the three variables $x=\frac{p^{+}}{P^{+}}$, $\xi=-\frac{\Delta^{+}}{2 P^{+}}$ and $t=-Q^{2}=\Delta^{2}$; where the light-cone coordinates are defined as $x^{\pm}=\frac{1}{\sqrt{2}}(x^{0}\pm x^{3}), \mathbf{x}_{T}=(x^{1},x^{2})$. We choose the light-front gauge $A^{+}=0$, so that the gauge link appearing in between the quark fields in Eqs.(\ref{gpd1}-\ref{gpd3}) becomes unity.

	\subsection{GPDs in impact parameter space}
The GPDs with zero skewness ($\xi =0$) in impact parameter space are  important for various  reasons: (i)  the density interpretation of the GPDs holds only in the impact parameter space for zero skewness \cite{PhysRevD.62.071503}, (ii)  the intuitive picture for various transverse SSAs in semi inclusive processes is based on the impact parameter representation of GPD $E^{q}$ \cite{urkardt:2003uw, burkardt2002impact}
 (iii)  It gives an intuitive connection between the Sivers asymmetry and the quark orbital angular momentum in certain models, (iv), the impact parameter representation allows to make analogies between chiral-odd quark GPDs and TMDs \cite{diehl2005spin}.  The parton correlator in impact parameter space is given by the Fourier transform as 
	\begin{eqnarray} \label{igpd}
		\mathcal{F}(x,\mathbf{b}_{T}; S)=\int\frac{d^2 \mathbf{\Delta}_{T}}{(2\pi)^2}\exp(-i \mathbf{\Delta}_{T}.\mathbf{b}_{T}) F(x,\mathbf{\Delta}_{T}; S).
	\end{eqnarray}
	Here $S$ denotes the polarization of the target.  The impact parameter $\mathbf{b}_{T}$ is conjugate to the transverse part of the momentum transfer $\mathbf{\Delta}_{T}$. 
	 The correlator defining the GPDs in impact parameter space is written as 
	\begin{eqnarray} \label{IGPD}
		\mathcal{F}^{q[\Gamma]}\left(x, \mathbf{b}_{T} ; S\right)
		=\frac{1}{2} \int \frac{d z^{-}}{2 \pi} e^{i x P^{+} z^{-}}\left\langle P^{+}, \mathbf{0}_{T} ; S\right| \bar{\psi}\left(z_{1}\right) \Gamma \nonumber \\
		\quad \times \mathcal{W}_{GPD}\left(z_{1} ; z_{2}\right) \psi\left(z_{2}\right)\left|P^{+}, \mathbf{0}_{T} ; S\right\rangle,
	\end{eqnarray}
	with $z_{1 / 2}=\left(0^{+}, \mp \frac{1}{2} z^{-}, \mathbf{b}_{T}\right)$.  Although the GPDs are expressed as off-forward matrix element and do not have a density interpretation, in the  impact parameter representation we can obtain diagonal matrix elements  and thus impact parameter dependent pdfs have a probabilistic interpretation.  The impact parameter dependent parton distributions (IPDpdfs) are given by 
		
\begin{eqnarray}		
		\mathcal{X}\left(x, \mathbf{b}_{T}^{2}\right)=\int \frac{d^{2} \mathbf{\Delta}_{T}}{(2 \pi)^{2}} e^{-i \mathbf{\Delta}_{T} \cdot \mathbf{b}_{T}} X\left(x, 0,-\mathbf{\Delta}_{T}^{2}\right).
	\end{eqnarray}
	where $X(x,0, -\mathbf{\Delta}_{T}^{2}) $ are the GPDs. The GPD correlators  (Eqs.(\ref{gpd1}-\ref{gpd3}) ) for different $\Gamma$ in the impact parameter space   are then obtained\cite{Meissner:2007rx} as 
	\begin{eqnarray} 
		\mathcal{F}^{q }\left(x, \mathbf{b}_{T} ; S\right)
		&=&  \mathcal{F}^{q [\gamma^{+}]}\left(x, \mathbf{b}_{T} ; S\right)
		=\mathcal{H}^{q}\left(x, \mathbf{b}_{T}^{2}\right)+\frac{\epsilon_{T}^{i j} b_{T}^{i} S_{T}^{j}}{M}\left(\mathcal{E}^{q}\left(x, \mathbf{b}_{T}^{2}\right)\right)^{\prime} \label{IGPD1}\\
		\tilde{\mathcal{F}}^{q }\left(x, \mathbf{b}_{T} ; S\right) &=& \tilde{\mathcal{F}}^{q [\gamma^{+} \gamma_{5}]}\left(x, \mathbf{b}_{T} ; S\right)
		=\lambda \tilde{\mathcal{H}}^{q}\left(x, \mathbf{b}_{T}^{2}\right), \label{IGPD2} \\
		\mathcal{F}_{T}^{q, j}\left(x, \mathbf{b}_{T} ; S\right) &=&\mathcal{F}^{q\left[i \sigma^{j+} \gamma_{5}\right]}\left(x, \mathbf{b}_{T} ; S\right)
		=\frac{\epsilon_{T}^{i j} b_{T}^{i}}{M}\left(\mathcal{E}_{T}^{q}\left(x, \mathbf{b}_{T}^{2}\right)+2 \tilde{\mathcal{H}}_{T}^{q}\left(x, \mathbf{b}_{T}^{2}\right)\right)^{\prime} \nonumber\\
		&+ & S_{T}^{j}\left(\mathcal{H}_{T}^{q}\left(x, \mathbf{b}_{T}^{2}\right)-\frac{\mathbf{b}_{T}^{2}}{M^{2}} \Delta_{b}
		 \tilde{\mathcal{H}}_{T}^{q}\left(x, \mathbf{b}_{T}^{2}\right)\right) \nonumber \\
		&+ & \frac{2 b_{T}^{j} \mathbf{b}_{T} \cdot \mathbf{S}_{T}-S_{T}^{j} \mathbf{b}_{T}^{2}}{M^{2}}\left(\tilde{\mathcal{H}}_{T}^{q}\left(x, \mathbf{b}_{T}^{2}\right)\right)^{\prime \prime}.\label{IGPD3}
	\end{eqnarray}
	Where 
	\begin{eqnarray} 
		\left(\mathcal{X}\left(x, \mathbf{b}_{T}^{2}\right)\right)^{\prime} &=& \frac{\partial}{\partial \mathbf{b}_{T}^{2}}\left(\mathcal{X}\left(x, \mathbf{b}_{T}^{2}\right) \right), \label{FTGPD} \\
		\Delta_b  \mathcal{X}\left(x, \mathbf{b}_{T}^{2}\right) &=&\frac{1}{{\mathbf{b}_T}^2} \frac{\partial}{\partial {\mathbf{b}_T}^2} \big[~{\mathbf{b}_T}^2  \frac{\partial}{\partial {\mathbf{b}_T}^2} \mathcal{X}\left(x, \mathbf{b}_{T}^{2}\right) \big].
	\end{eqnarray}
	
 Eq.(\ref{IGPD1})  describes the distribution of unpolarized quarks
	carrying the longitudinal momentum fraction $x$ at a transverse position $b_{T}$. For a transversely polarized target, this distribution has a spin independent part given by $\mathcal{H}$ and a spin dependent part  proportional to the first order derivative of $\mathcal{E}$.  Due to spin dependent term,     the  distribution in impact parameter space is not axially symmetric as it depends on the direction of the impact parameter $\mathbf{b}_{T}$, i.e,  the spin-dependent term causes a distortion of the distribution.  
 This distortion effect  can be quantified through the flavor dipole moment \cite{burkardt2005transverse}:
	\begin{eqnarray} \label{dipolemom}
		d^{q, i}=\int d x \int d^{2} \mathbf{b}_{T} b_{T}^{i} \mathcal{F}^{q}\left(x, \mathbf{b}_{T} ; S\right)=-\frac{\epsilon_{T}^{i j} S_{T}^{j}}{2 M} \int d x E^{q}(x, 0,0)=-\frac{\epsilon_{T}^{i j} S_{T}^{j}}{2 M}  \kappa^{q}. 
	\end{eqnarray}
	Here $\kappa^{q}$ is the contribution of the quark flavor $q$  to the anomalous magnetic moment of the nucleon. The flavor dipole moments for the light quarks in the nucleon are therefore of the order  $0.2 ~fm$ which is quite significant in comparison to the size
	of the nucleon.  Similarly, in Eq.(\ref{IGPD3}) there are two terms which   generate  distortions, one is determined by the first derivative of $\mathcal{E}_{T}+2\tilde{\mathcal{H}}_{T}$ and the other is given by the second derivative of $\tilde{\mathcal{H}}_{T}$. We will see later that the specific form of the relations between GPDs and TMDs depends on the number of the derivatives of the GPDs in impact parameter space.

	\subsection{Transverse momentum dependent	parton distributions (TMDs)}
		
	The quark TMDs are defined through the unintegrated quark-quark correlator for semi-inclusive deep inelastic scattering
	(SIDIS) \cite{mulders1996complete,bacchetta2007semi}. The TMDs depend on the longitudinal momentum fraction $x$ 
	of the active quark and the quark transverse momentum $\mathbf{p}_{T}$. The TMDs provide a three-dimensional view of the parton distributions in momentum space.
	In a hadronic state $|P,S\rangle$ with momentum P and polarization S, the TMDs can be defined through the quark-quark correlation function as
	\begin{eqnarray}
		\Phi^{q[\Gamma]}\left(x, \mathbf{p}_{T}; \mathbf{S}\right) 
		&=& \frac{1}{2} \int \frac{d z^{-}}{2 \pi}\frac{d^{2} \mathbf{z}_{T}}{(2 \pi)^{2}} e^{i k \cdot z}\left\langle P; S \right| \bar{\psi}\left(-\frac{1}{2} z\right) \Gamma
		 \nonumber \\
		&& \times\left.\mathcal{W}_{TMD}\left(-\frac{1}{2} z ; \frac{1}{2} z\right) \psi\left(\frac{1}{2} z\right)|P ; S \rangle\right|_{z^{+}=0^{+}}, \label{tmd}
	\end{eqnarray}	
	in which a summation over the colour of the quark fields is implicit.
	In the chosen frame,  the nucleon four momentum $P\equiv \left(P^{+},\frac{M^{2}}{P^{+}},\mathbf{0}\right)$, and virtual photon momentum $q\equiv \left(x_{B}P^{+}, \frac{Q^{2}}{x_{B}P^{+}},\mathbf{0}\right)$, where $x_{B}=\frac{Q^{2}}{2P.q}$ is the Bjorken variable and $Q^{2}=-q^{2}$.  
	The covariant spin vector S for the nucleon with helicity $\lambda$ is defined as  $(S^{+}=\frac{\lambda P^{+}}{M}, S^{-}=-\frac{\lambda P^{-}}{M},     \mathbf{S}_{T})$. The Wilson line in TMD correlators is far more complicated than that in the GPD correlators and is given by \cite{Bacchetta:2008af}
	\begin{eqnarray}
		&& \left.\mathcal{W}_{TMD}\left(-\frac{1}{2} z ; \frac{1}{2} z\right) \right|_{z^{+}=0^{+}} 
		=  \left[0^{+},-\frac{1}{2} z^{-},-\frac{1}{2} \mathbf{z}_{T} ; 0^{+},+\infty^{-},-\frac{1}{2} \mathbf{z}_{T}\right] \times \nonumber\\
		&&~~~~~~~~~~~~\left[0^{+},+\infty^{-},-\frac{1}{2} \mathbf{z}_{T}  ;  0^{+},+\infty^{-}, \frac{1}{2} \mathbf{z}_{T}\right] 
 \left[0^{+},+\infty^{-}, \frac{1}{2} \mathbf{z}_{T} ; 0^{+}, \frac{1}{2} z^{-}, \frac{1}{2} \mathbf{z}_{T}\right],
	\end{eqnarray}
	where the future pointing Wilson line is running along the positive  $z^{-}$ direction to $+\infty$ for SIDIS, while in the Drell-Yan process the Wilson lines runs along the opposite direction towards $-\infty$ \cite{Collins:2002kn}. 	The leading twist quark TMDs are obtained from correlator in (\ref{tmd}) by using the same three $\Gamma$  matrices($\gamma^{+}, \gamma^{+}\gamma_{5}, \sigma^{+j}\gamma_{5}$). The corresponding antiquark TMDs can be obtained by the same light cone correlation functions  by using charge conjugated fields. There exits eight leading twist quark TMDs, which are all real valued.  Taking $\Gamma=\gamma^+$ in Eq.(\ref{tmd}), we get
	\begin{eqnarray}
		\Phi^{q}\left(x, \mathbf{p}_{T} ; S\right) &=& \Phi^{q\left[\gamma^{+}\right]}\left(x, \mathbf{p}_{T} ; S\right) \nonumber\\
	&=& f_{1}^{q}\left(x, \mathbf{p}_{T}^{2}\right)-\frac{\epsilon_{T}^{i j} p_{T}^{i} S_{T}^{j}}{M} f_{1 T}^{\perp q}\left(x, \mathbf{p}_{T}^{2}\right).\label{tmd1}
	\end{eqnarray}
  $f_{1}^{q}$ is the unpolarized quark distribution for a quark flavor q, $f_{1 T}^{\perp q}$ represent the (naive) time reversal odd (T-odd) Sivers function which appears for a transversely polarized target ($S_{T}\neq0$).  
	The Sivers function describes the distribution of the unpolarized quark carrying  a longitudinal momentum fraction $x$ and transverse momentum $\mathbf{p}_{T}$ in a transversely polarized target. If the second term on the right-hand side in (\ref{tmd1}) is nonzero then the quark TMD correlator $\Phi^{q}$ is not axially symmetric in the $\mathbf{p}_{T}$ plane i.e.,  the distribution  becomes distorted. This distortion is
	supposed to be  the origin of various observed single spin asymmetries in hard semi-inclusive reactions \cite{barone2002drago,anselmino1996polarized}. Setting $\Gamma=\gamma^+\gamma_5$ and $i\sigma^{j+}\gamma_5$ in Eq.(\ref{tmd}), we have
	\begin{eqnarray} 
		\tilde{\Phi}^{q}\left(x, \mathbf{p}_{T} ; S\right) &=& \Phi^{q\left [\gamma^{+} \gamma_{5}\right]}\left(x, \mathbf{p}_{T} ; S\right) 
		=\lambda g_{1 L}^{q}\left(x, \mathbf{p}_{T}^{2}\right)+\frac{\mathbf{p}_{T} \cdot \mathbf{S}_{T}}{M} g_{1 T}^{q}\left(x, \mathbf{p}_{T}^{2}\right);\label{tmd2}\\
		\Phi_{T}^{q, j}\left(x, \mathbf{p}_{T} ; S\right) &=& \Phi^{q\left[i \sigma^{j+} \gamma_{5}\right]}\left(x,\mathbf{p}_{T} ; S\right)
		=-\frac{\epsilon_{T}^{i j} \mathbf{p}_{T}^{i}}{M} h_{1}^{\perp q}\left(x, \mathbf{p}_{T}^{2}\right)+\frac{\lambda \mathbf{p}_{T}^{j}}{M} h_{1 L}^{\perp q}\left(x, \mathbf{p}_{T}^{2}\right) \nonumber \\
		+& S_{T}^{j} &\!\!\!  \left(h_{1 T}^{q}\left(x, \mathbf{p}_{T}^{2}\right)
		+  \frac{\mathbf{p}_{T}^{2}}{2 M^{2}} h_{1 T}^{\perp q}\left(x, \mathbf{p}_{T}^{2}\right)\right)
		+\frac{2 \mathbf{p}_{T}^{j} \mathbf{p}_T \cdot \mathbf{S}_{T}-S_{T}^{j} \mathbf{p}_T^{2}}{2 M^{2}} h_{1 T}^{\perp q}\left(x, \mathbf{p}_T^{2}\right). \label{tmd3}
	\end{eqnarray}
	Here $h_1^{\perp q}$ is the Boer-Mulders function, which gives the distribution of transversely polarized quarks inside an unpolarized hadron. The process dependence of the Wilson line leads to a sign difference in the T-odd TMDs, e.g. $f_{1 T}^{\perp q}|_{SIDIS}=- f_{1 T}^{\perp q}|_{DY}$. Whereas in hadron-hadron scattering with hadronic final states, even more complicated paths for the Wilson line can  arise \cite{bomhof2004gauge,bomhof2007gluonic,Buffing:2012sz,Buffing:2013kca}.
 There are
	altogether six T-even TMDs and two (Sivers and Boer Mulders) T-odd TMDs. The $p_T$ integrated function of $f_{1}^{\nu}(x,p_T^{2})$ gives the unpolarized parton distribution $f_{1}^{\nu}(x)$ and $g_{1 L}^{\nu}(x,p_T^{2})$ gives the helicity distribution $g_{1 L}^{\nu}(x)$.  The transversity TMD $h_{1}^{\nu}(x,p_T^{2})$  defined as 
	\begin{eqnarray}
		h_{1}^{\nu}\left(x, \mathbf{p}_T^{2}\right)=h_{1 T}^{\nu}\left(x, \mathbf{p}_T^{2}\right)+\frac{\mathbf{p}_T^{2}}{2 M^{2}} h_{1 T}^{\perp \nu}\left(x,\mathbf{p}_T^{2}\right),
	\end{eqnarray}
	  when integrated over the transverse momentum gives the transversity parton distribution  $h_{1}^{\nu}(x)$. 
	  
Like the $p_T$ integrated TMDs, some GPDs in the limit  $\xi=t=0$,  are also related to the  twist-2 PDFs.  Thus we  get some model independent relations between GPDs and TMDs:
	\begin{eqnarray} 
		f_{1}^{q}(x)&=&\int d^{2} \mathbf{p}_T f_{1}^{q}\left(x, \mathbf{p}_T^{2}\right)
		= H^{q}(x,0,0)=\int d^{2} \mathbf{b}_{T} \mathcal{H}^{q / g}\left(x, \mathbf{b}_{T}^{2}\right), \label{unpolarized}\\
		g_{1}^{q}(x) &=& \int d^{2} \mathbf{p}_T g_{1 L}^{q}\left(x, \mathbf{p}_T^{2}\right)
		=\tilde{H}^{q}(x, 0,0)=\int d^{2} \mathbf{b}_{T} \tilde{\mathcal{H}}^{q}\left(x, \mathbf{b}_{T}^{2}\right), \label{helicity}\\
		h_{1}^{q}(x) &=& \int d^{2} \mathbf{p}_T \left(h_{1 T}^{q}\left(x, \mathbf{p}_T^{2}\right)+\frac{\mathbf{p}_T^{2}}{2 M^{2}} h_{1 T}^{\perp q}\left(x, \mathbf{p}_T^{2}\right)\right) 
		=H_{T}^{q}(x, 0,0) \nonumber \\
		&=& \int d^{2} \mathbf{b}_{T}\left(\mathcal{H}_{T}^{q}\left(x, \mathbf{b}_{T}^{2}\right)-\frac{\mathbf{b}_{T}^{2}}{M^{2}} \Delta_{b} \tilde{\mathcal{H}}_{T}^{q}\left(x, \mathbf{b}_{T}^{2}\right)\right). \label{transversity}
	\end{eqnarray}
	
	These relations give important constraints on models, and are satisfied in our model. 
Next, we investigate a few nontrivial model dependent relations.  
	
	\section{Model dependent relations between GPDs and TMDs }\label{relations1}

In \cite{Meissner:2007rx} all possible model dependent relations between GPDs and TMDs are systematically studied. The model independent relations as given in the previous section  are called relations of the first kind.  There are a few relations in momentum space. There are also model dependent relations that connect the GPDs to the IPDpdfs or their derivatives in $b_T$ space. 
These relations are called second, third and fourth type depending on the number of derivatives present in the relation.   In this section,  we investigate  nontrivial relations between the TMDs and GPDs, as well as some relations between the different TMDs in  our model. We  also compare LFQDQ model results with the results of two other spectator models, namely, the scalar diquark spectator model of the nucleons \cite{brodsky2002final} and  a quark target model treated in perturbative QCD \cite{Meissner:2007rx}.\\
 Incorporating the FSI effect into the wavefunctions \cite{Chakrabarti:2019wjx}, the Sivers  $f_{1 T}^{\perp q}$ and the Boer- Mulders $h_{1}^{\perp q}$ functions can be written as
		\begin{eqnarray}
		f_{1T}^{\perp q}(x,\mathbf{p}_T^2) &=&\Big{(}C^2_{S} N_{S}^{\nu 2}-\frac{1}{3}C_{A}^2 N_{0}^{{\nu}2}\Big{)} f^{\nu}(x,\mathbf{p}_T^{2})   \label{Siverstmd}\\
		h_{1}^{\perp q}(x,\mathbf{p}_T^2) &=& \Big{(}C^2_{S} N_{S}^{\nu 2}+(\frac{1}{3} N_{0}^{{\nu}2}+\frac{2}{3}N_{1}^{{\nu}2})C_{A}^2\Big{)}f^{\nu}(x,\mathbf{p}_T^{2}), \label{BMtmd}
	\end{eqnarray}
	where 
	\begin{eqnarray}
		f^{\nu}(x,\mathbf{p}_T^{2})=(-C_{F}\alpha_{s}) \times \frac{1}{x}\Big{[}\frac{\textbf{p}_T^2+B(x)}{\textbf{p}_T^2}\Big{]}\log\Big{[}\frac{\textbf{p}_T^2+B(x)}{B(x)}\Big{]}  \nonumber \\
		\times  \frac{1}{16\pi^3}A_{1}^{\nu}(x)A_{2}^{\nu}(x)\exp\Big{[}-a(x)\textbf{p}_T^2\Big{]},
	\end{eqnarray} 
	where $B(x)$ and $a(x)$ are defined in Eq.(\ref{Bx}) and Eq.(\ref{ax}) respectively. In the final state interactions, gluon exchange strength $\frac{e_{1}e_{2}}{4\pi} \rightarrow-C_{F} \alpha_{s}$. 
While the pretzelosity $h_{1 T}^{\perp q}$ TMDs can be written as \cite{Maji:2017bcz} 
	\begin{equation} \label{pretzelositytmd}
		h_{1 T}^{\nu \perp}\left(x, \mathbf{p}_T^{2}\right)=-\left(C_{S}^{2} N_{S}^{\nu 2}-C_{V}^{2} \frac{1}{3} N_{0}^{\nu 2}\right) \frac{2 \ln (1 / x)}{\pi \kappa^{2}} x^{2a_{2}^{\nu}-2}(1-x)^{2b_{2}^{\nu}-1} \exp \left[-a(x) \mathbf{p}_T^{2}\right],
	\end{equation}
	 and found to satisfy the inequality relation\cite{Maji:2017wwd}
	\begin{eqnarray}
		|h_{1}^{\perp} (x,\mathbf{p}_T^{2})|>|f_{1 T}^{\perp}(x,\mathbf{p}_T^{2})|.
	\end{eqnarray}
	From Eq.(\ref{Siverstmd}) and Eq.(\ref{BMtmd}), we can easily check that Boer-Mulders function is proportional to the Sivers function. The Boer-Mulders function can be parametrized \cite{barone2010boer, Maji:2017wwd} as 
	\begin{eqnarray}
		h_{1}^{\perp \nu}\left(x, \mathbf{p}_T^{2}\right) = \lambda^{\nu} f_{1 T}^{\perp \nu}\left(x, \mathbf{p}_T^{2}\right), \label{BM-S}
	\end{eqnarray}
	where  
	\begin{eqnarray} \label{lambdanu}
\lambda^{\nu}=\frac{\Big{(}C^2_{S} N_{S}^{\nu 2}+(\frac{1}{3} N_{0}^{{\nu}2}+\frac{2}{3}N_{1}^{{\nu}2})C_{A}^2\Big{)}}{\Big{(}C^2_{S} N_{S}^{\nu 2}-\frac{1}{3}C_{A}^2 N_{0}^{{\nu}2}\Big{)}}.
	\end{eqnarray}
 Putting in the model parameters in Eq.(\ref{lambdanu}),  we  get $\lambda^{u}=2.29$ and $\lambda^{d}=-1.08$. Since the Sivers function is negative for up quarks and positive for the down quarks. By the above expression we can conclude that the Boer-Mulders function is negative for both up and down quarks.
	\subsection{ Moment relations between GPDs and TMDs in momentum space}
The $n^{th}$ moment of the GPD $X$ is  defined as,
	\begin{eqnarray}\label{gpdsmom}
		X^{(n)}(x)
		=\frac{1}{2 M^{2}} \int d^{2} \mathbf{\Delta}_{T}\left(\frac{\mathbf{\Delta}_{T}^{2}}{2 M^{2}}\right)^{n-1} X\left(x, 0,-\frac{\mathbf{\Delta}_{T}^{2}}{(1-x)^{2}}\right)
	\end{eqnarray} 
	and similarly the  moment of the TMD Y is defined as
	\begin{eqnarray}\label{tmdsmom}
		Y^{(n)}(x)=\int d^{2} \textbf{p}_T\left(\frac{\textbf{p}_T^{2}}{2 M^{2}}\right)^{n} Y\left(x, \textbf{p}_T^{2}\right) 
	\end{eqnarray} 
 The  moments for  GPD $E^{q}$ and $(E_{T}^{q}+2 \tilde{H}_{T}^{q})$  are obtained as
	\begin{eqnarray} \label{gpdeqn}
		E^{q (n)}(x)=-\left(C_{s}^{2} N_{s}^{2}-\frac{1}{3}C_{A}^{2} N_{0}^{\nu 2}\right) \pi  2^{n+1} \left(\frac{1}{M^2}\right)^n \Gamma (n)  \nonumber \\
		x^{a_{1}^{\nu}+a_{2}^{\nu}-1} (1-x)^{b_{1}^{\nu}+b_{2}^{\nu}+2} \left(\frac{\log (1/x)}{\kappa ^2 (x-1)^4}\right)^{-n},
	\end{eqnarray}
	and
	\begin{eqnarray} \label{gpdEp2Hqn}
		E_{T}^{q (n)}(x)+2 \tilde{H}_{T}^{q (n)}(x)&=&\left( C_{s}^{2} N_{s}^{2}+ C_{A}^{2} \left(\frac{1}{3}N_{0}^{\nu 2}+\frac{2}{3} N_{1}^{\nu 2}\right)\right) \pi  2^n \left(\frac{1}{M^2}\right)^n \Gamma (n)  \nonumber \\
		&&x^{a_{1}^{\nu}+a_{2}^{\nu}-1} (1-x)^{b_{1}^{\nu}+b_{2}^{\nu}+2} \left(\frac{  \log (1/x)}{\kappa ^2 (x-1)^4}\right)^{-n}.
	\end{eqnarray}
The analytic form of the  moments for Sivers and Boer-Mulders functions  are too complicated and lengthy to present here. The moment relation between Sivers function and GPD $E^q$  in scalar diquark model  was  obtained in  \cite{Meissner:2007rx}  and has the general form 
	\begin{eqnarray} \label{sdmstmd}
		f_{1 T}^{\perp q(n)}(x)
		&=&-\frac{g^{2} e_{q} e_{s}(1-x)}{16(2 \pi)^{2}} \frac{\left(m_{q}+x M\right) \tilde{M}^{2 n-2}(x) H_{-n}}{2^{n} M^{2 n-1} \sin (n \pi)} \nonumber \\
		&= & -\frac{e_{q} e_{s}}{2(2 \pi)^{2}(1-x)} \frac{H_{-n} \Gamma(2-2 n)}{\Gamma^{2}(1-n)} E^{q(n)}(x).
	\end{eqnarray}
	 The relation given by Eq. (\ref{sdmstmd}) generally holds for $0\leq n \leq 1$, i.e., $n$  is not necessarily an integer where $H_n$ is the analytic continuation of the harmonic number for non-integer $n$.  In  LFQDQ model,  it is not possible to obtain a similar relation analytically for a general value of $n$, however,  relations similar to Eq.  (\ref{sdmstmd}) can be derived  in the scalar diquark model for three particular values of $n$ as
	\begin{eqnarray} 
		f_{1 T}^{\perp q(0)}(x)  &= &\frac{\pi e_{q} e_{s}}{48(1-x)} E^{q}(x, 0,0), \label{sdmstmd0}\\
		f_{1 T}^{\perp q(1 / 2)}(x) & =& \frac{2  e_{q} e_{s} \ln( 2)} {(2 \pi)^{3}(1-x)} E^{q(1 / 2)}(x) , \label{sdmstmd05}\\
		f_{1 T}^{\perp q(1)}(x) &=&  \frac{e_{q} e_{s}}{4(2 \pi)^{2}(1-x)} E^{q(1)}(x), \label{sdmstmd1}
	\end{eqnarray}
	 in agreement with \cite{Meissner:2007rx}.
	\begin{figure} 
		\centering
		\includegraphics[scale=0.32]{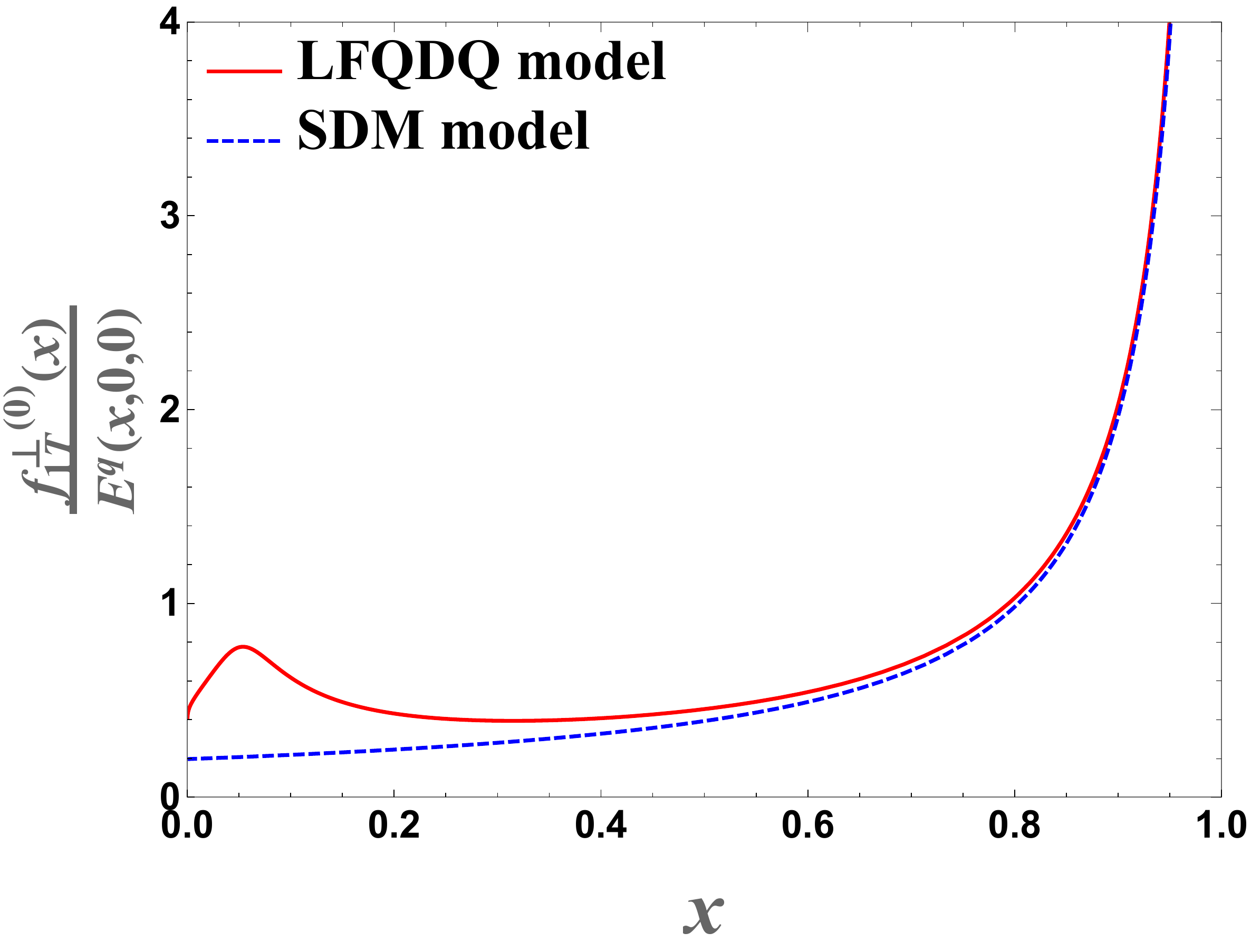}
		\includegraphics[scale=0.32]{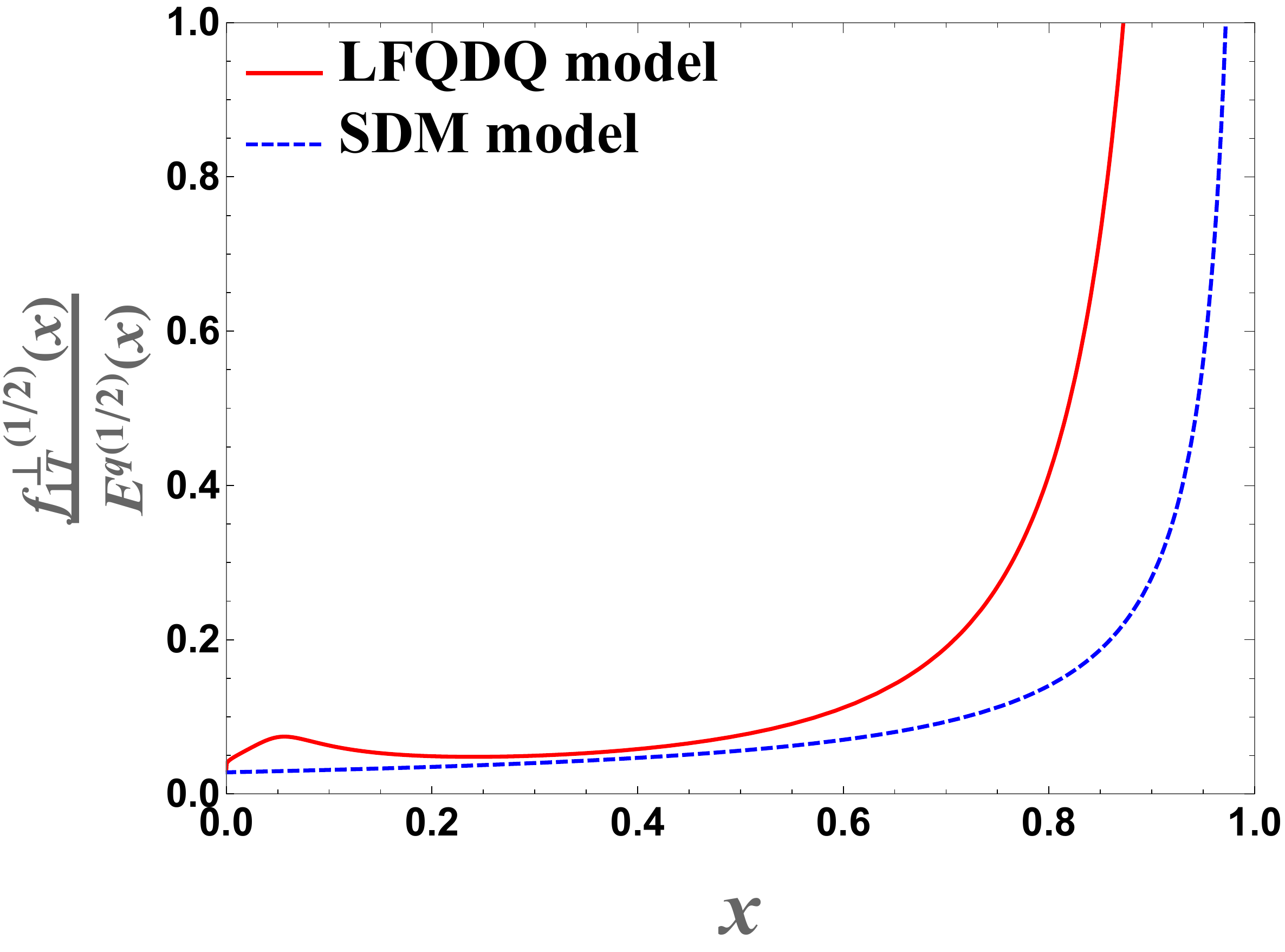}
		\includegraphics[scale=0.32]{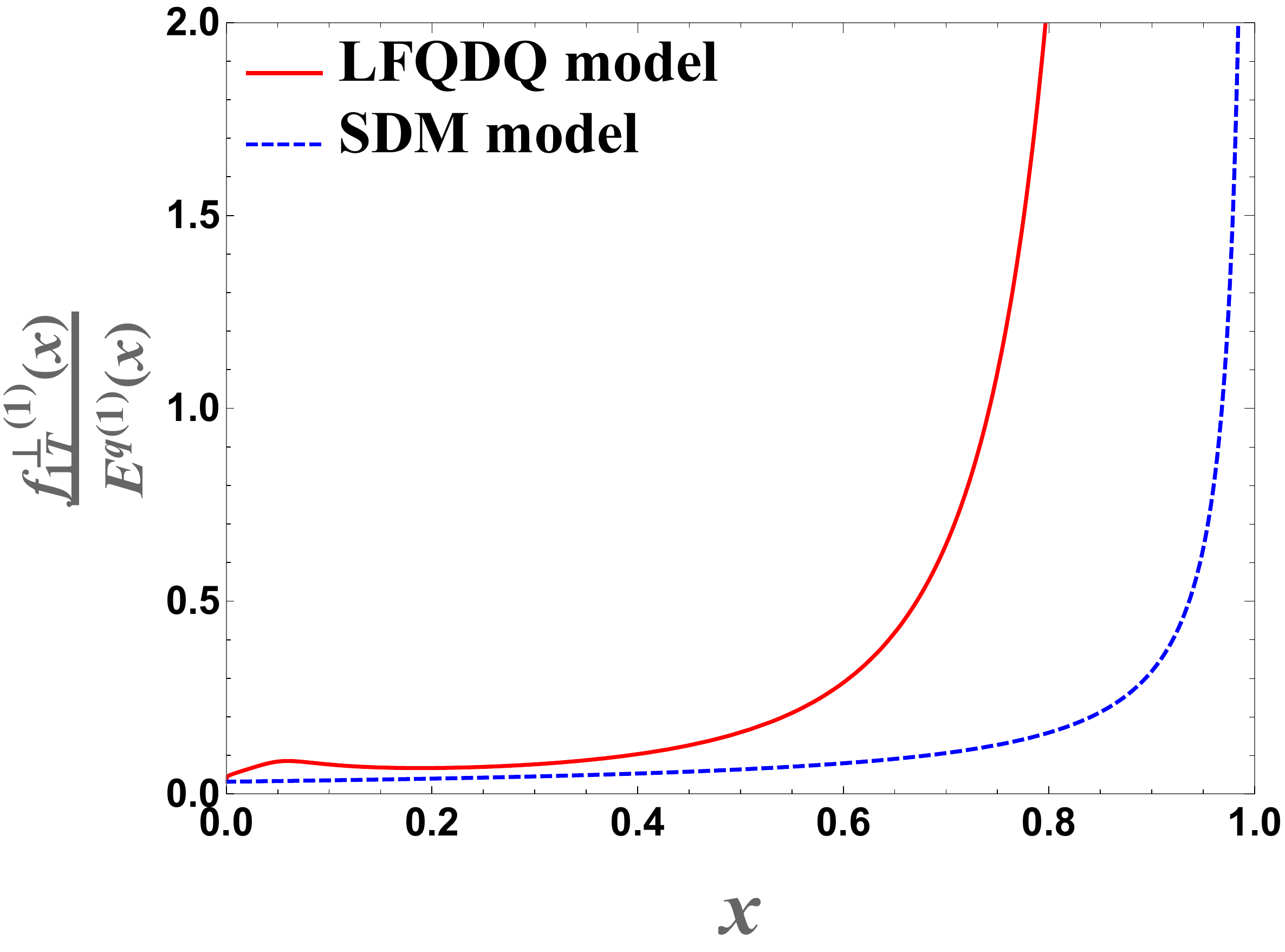}
		\caption{The left plot is corresponding the ratio of the zeroth moment of Sivers TMD for the up quarks $f_{1T}^{\perp u(0)}$(x) to the corresponding GPD $E^{u}(x,0,0)$. The right one is for the ratio of the $f_{1 T}^{\perp u(1/2)}(x)$ to $E^{u (1/2)}(x)$, and  the lower one is for the ratio of the $f_{1 T}^{\perp u(1)}(x)$ to $E^{u (1)}(x).$}
		\label{figstmd}
	\end{figure}
\begin{figure} 
	\centering
	\includegraphics[scale=0.32]{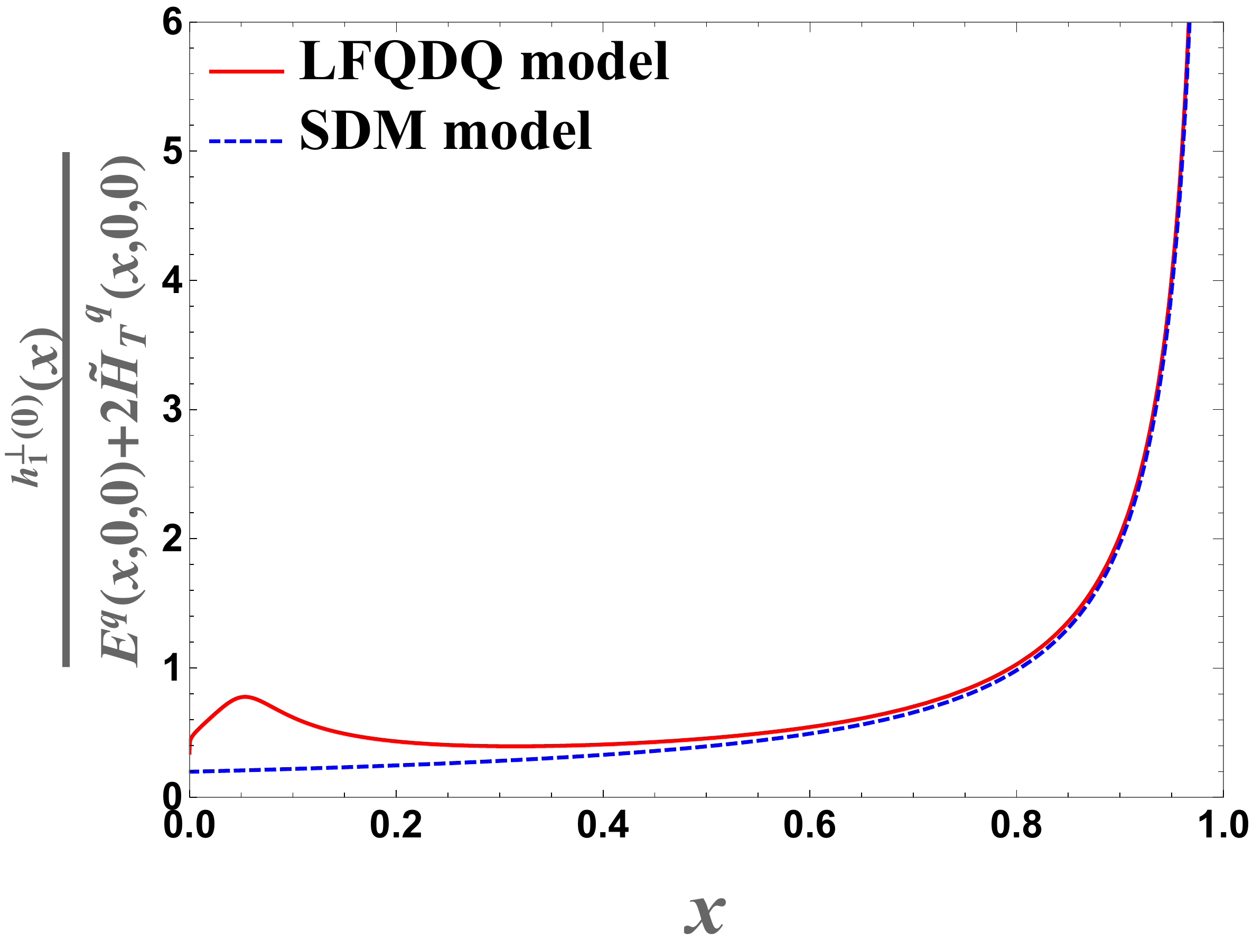}
	\includegraphics[scale=0.32]{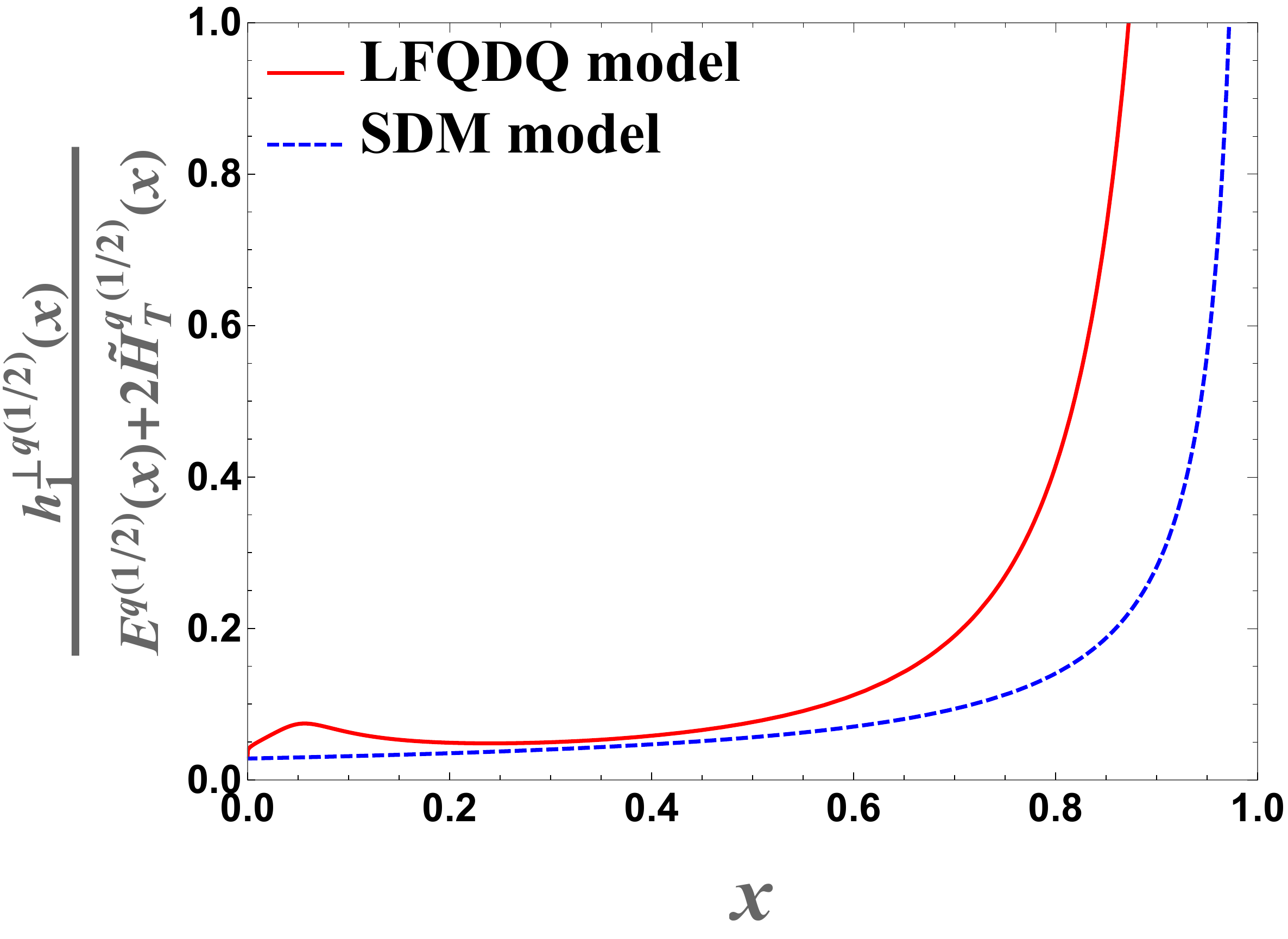}
	\includegraphics[scale=0.32]{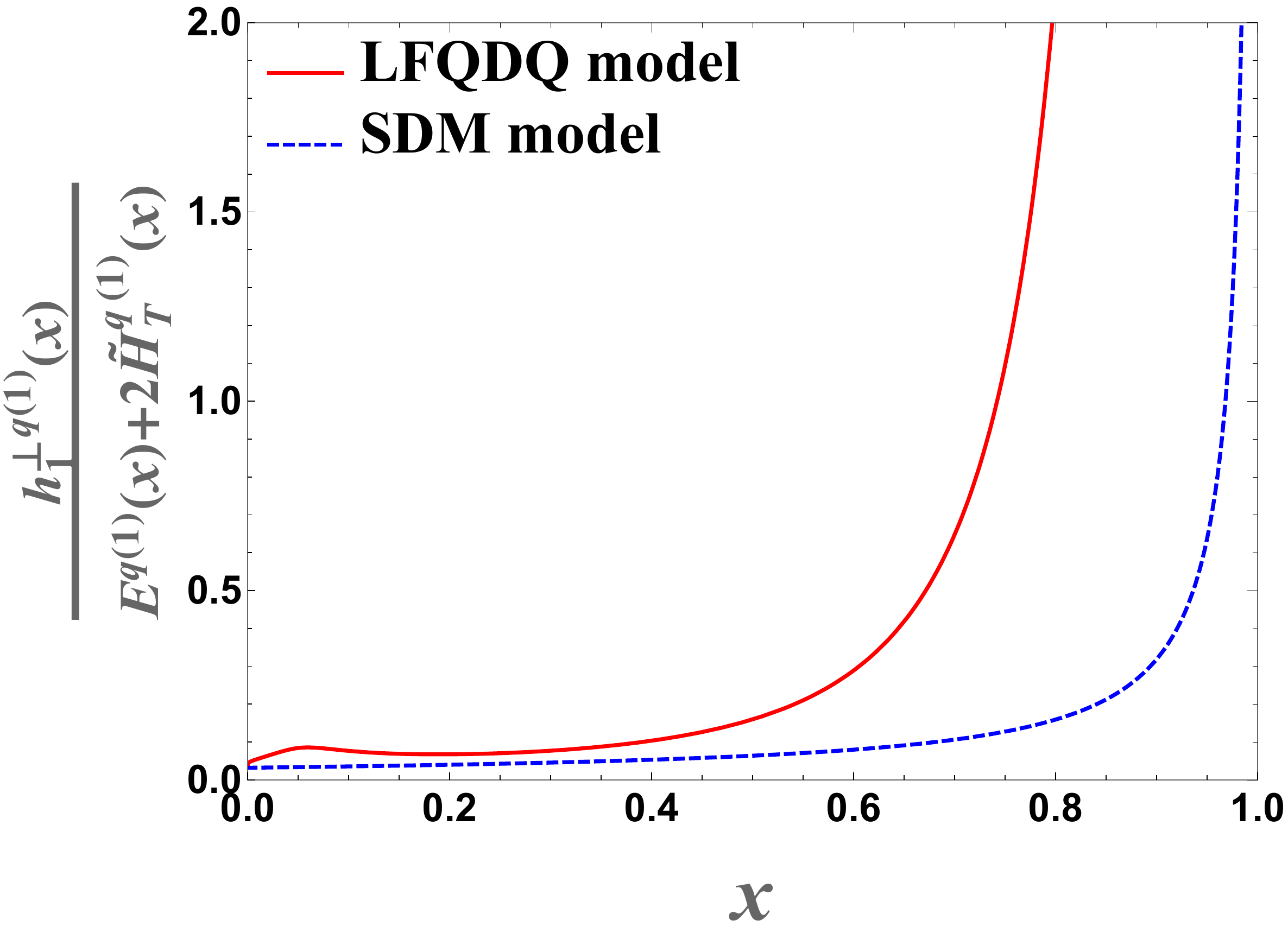}
	\caption{The left plot is corresponding the ratio of the zeroth moment of Boer-Mulders TMD for the up quarks $h_{1}^{\perp u(0)}$(x) to the corresponding GPD, $E_{T}^{u}(x,0,0)+2\tilde{H}_{T}^{u}(x,0,0)$. The right one is for the ratio of the $h_{1 T}^{\perp q(1/2)}(x)$ to $E^{u (1/2)}(x)+2\tilde{H}_{T}^{u(1/2)}(x)$, and  The lower one is for the ratio of the $h_{1 }^{\perp q(1)}(x)$ to $E^{u (1)}(x)+2\tilde{H}_{T}^{u(1)}(x).$}
	\label{figbmtmd}
\end{figure}
The moment relations for the Sivers function, $f_{1T}^{\perp q}$ and corresponding GPD, $E^{q}$ in the LFQDQ model are evaluated numerically and the results   for the up quarks are  compared with scalar diquark model results in Figure \ref{figstmd}. It is seen from the plot that the zeroth moment of the Sivers function in both models obey the same relation   for $x > 0.5$; although the qualitative behaviour for the next two moments $(n=1/2, 1)$ are similar, they do not match except for a narrow range roughly in the region $0.1 < x  < 0.5$.   Furthermore, in LFQDQ model,  a kink  appears at  around $x=0.05$. This is due to that fact that  both the functions $f_{1 T}^{\perp q}(x)$ and GPD $E^{q}(x)$ has maxima at same $x$. The other T-odd TMD i.e, Boer-Mulders function and the corresponding GPD also show similar behavior, which can be seen in Fig. \ref{figbmtmd}. 
	
	 In LFQDQ model, the relations between the pretzelosity, $h_{1 T}^{\perp q}$ and the GPD, $\tilde{H}_{T}^{q}$ can be given analytically. By using the Eqs. (\ref{gpdsmom}), and (\ref{tmdsmom}) the $n^{th}$ moment of the pretzelosity TMD, $h_{1 T}^{\perp q(n)}$ and the $n^{th}$ moment of the corresponding GPD $\tilde{H}_{T}^{q (n)}$ is given as
	\begin{eqnarray}
		h_{1 T}^{\perp q (n)}(x)&=&-\left(C_s^2 N_s^2- \frac{1}{3} C_A^2 N_0^{ \nu 2}\right)\frac{2^{1-n}}{\kappa ^2} \left(\frac{1}{M^2}\right)^n \Gamma (n+1) x^{2a_{2}^{\nu}-2}(1-x)^{2b_{2}^{\nu}-2} \nonumber\\
		&&~~~ \times \log \left(\frac{1}{x}\right) a(x)^{-n-1}  \label{pretn}\\
		\tilde{H}_{T}^{q (n)}(x)&=& \left( C_{s}^{2} N_{s}^{2}-\frac{1}{3}C_{A}^{2} N_{0}^{\nu 2}\right)\pi  2^n x^{2a_{2}^{\nu}-2} (1-x)^{2b_{2}^{\nu}+3} \left(\frac{1}{M^2}\right)^n \Gamma (n) a(x)^{-n}  \label{gpdHTn}
	\end{eqnarray}
	and they satisfy the relation
	\begin{equation}  \label{PretnandHTn}
		h_{1T}^{\perp q (n)}  =\frac{2^{1-2 n} \Gamma (n+1)}{\pi  (1-x)^2 \Gamma (n)} \tilde{H}_{T}^{q (n)}(x)
	\end{equation}
	which holds for $0\leq n \leq 1$.  Note that this relation is the same with or without the axial-vector diquark in the model.
	The explicit forms of the relation in Eq.(\ref{PretnandHTn}) for three different values of $n$ are as follows.
	\begin{eqnarray} 
		h_{1T}^{\perp q(0)}(x) &=& \frac{2}{(1-x)^2}\tilde{H}_{T}^q(x,0,0), \label{pret0}\\
		h_{1T}^{\perp q(1/2)}(x) &=& \frac{1}{2\pi(1-x)^2}\tilde{H}_{T}^{q(1/2)}(x), \label{pret05}\\
		h_{1T}^{\perp q(1)}(x) &=&	\frac{1}{2\pi(1-x)^2}\tilde{H}_{T}^{q(1)}(x). \label{pret1}
	\end{eqnarray}
	These relations can be compared with the corresponding relations in the scalar diquark model and quark target models , which are given by  \cite{Meissner:2007rx} 
	\begin{eqnarray} 
		h_{1T}^{\perp q(0)}(x) &=& \frac{3}{(1-x)^2}\tilde{H}_{T}^q(x,0,0), \label{sdmpret0}\\
		h_{1T}^{\perp q(1/2)}(x) &=&
		\frac{8}{(2\pi)^2(1-x)^2}\tilde{H}_{T}^{q(1/2)}(x), \label{sdmpret05} \\
		h_{1T}^{\perp q(1)}(x) &=&
		\frac{1}{2\pi(1-x)^2}\tilde{H}_{T}^{q(1)}(x). \label{sdmpret1}
	\end{eqnarray}
	Note that Eq.(\ref{pret1}) exactly matches with Eq. (\ref{sdmpret1}), and the other two relations (Eqs. \ref{pret0},\ref{pret05}) have similar structure as Eq.(\ref{sdmpret0},\ref{sdmpret05}), except the constant factors, which may be model dependent. The ratios of $h_{1T}^{\perp q(n)}(x)$ and $\tilde{H}_{T}^{q(n)}(x)$  for $0\le n \le1$ go as $1/(1-x)^2$ in the scalar diquark model considered in \cite{Meissner:2007rx} as well as in our model with or without the axial-vector diquark indicates that this property  is  true  in any quark-diquark model.
	
\begin{figure} 
\centering
\includegraphics[scale=0.32]{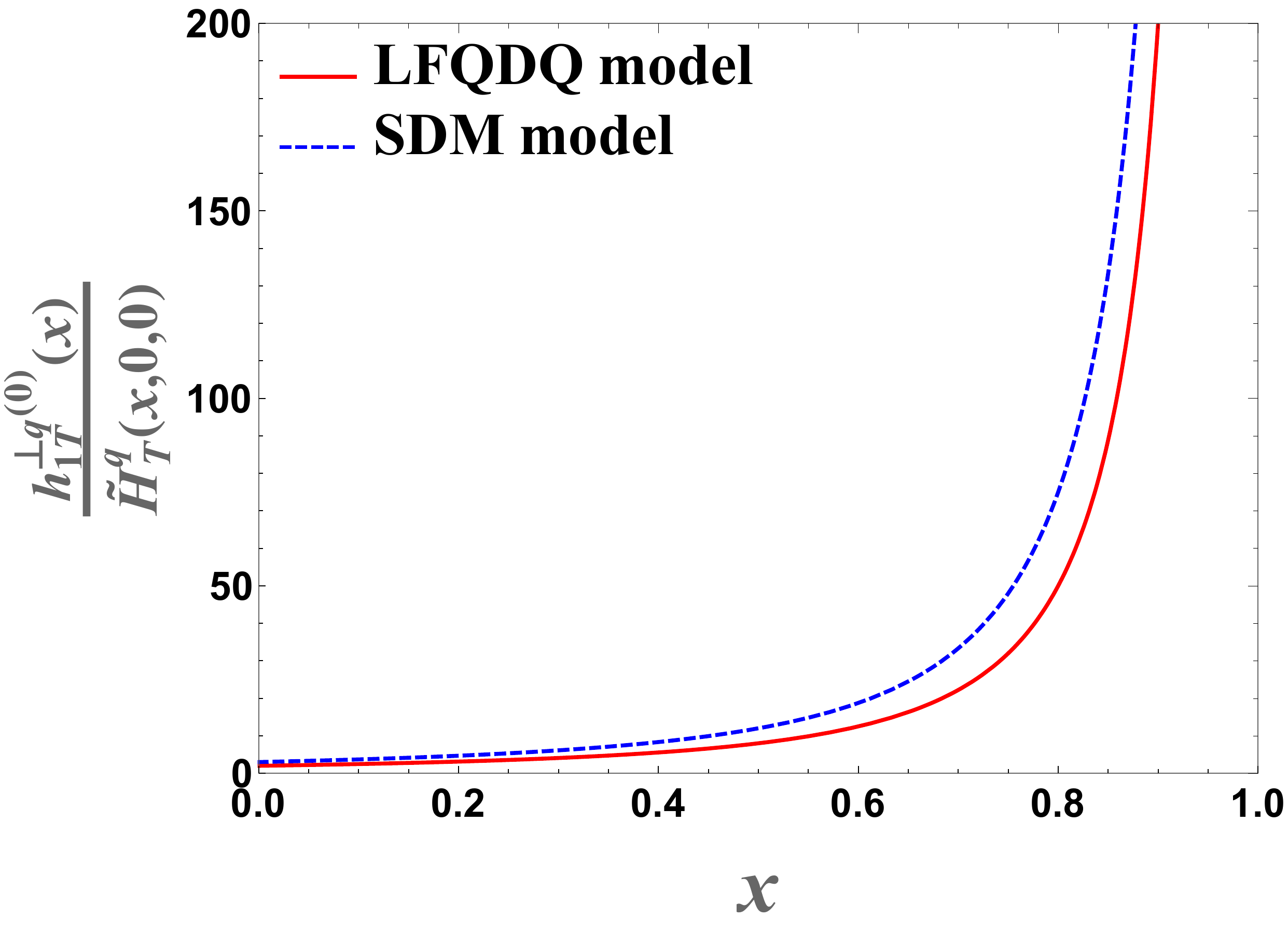}
\includegraphics[scale=0.32]{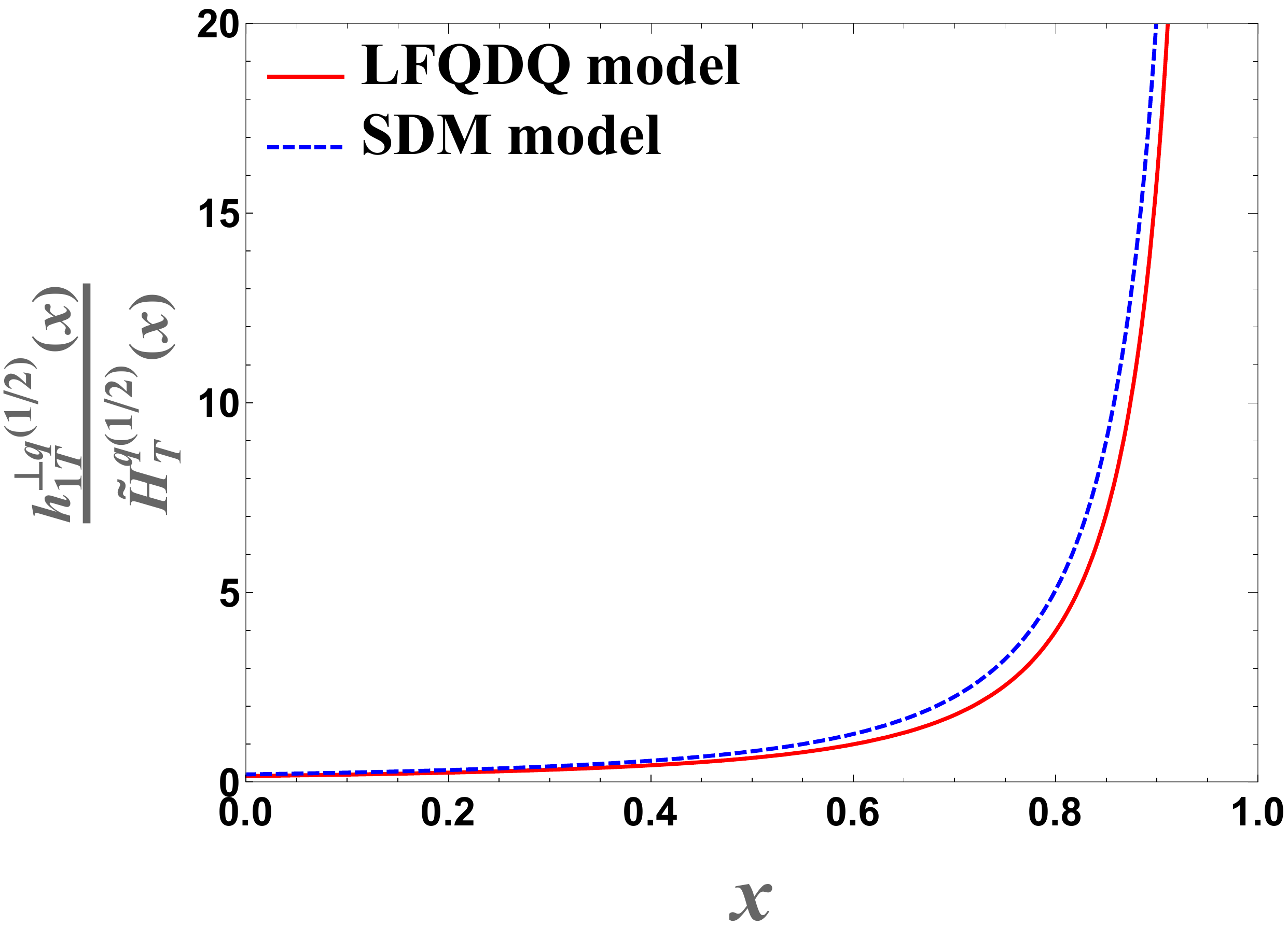}
\includegraphics[scale=0.32]{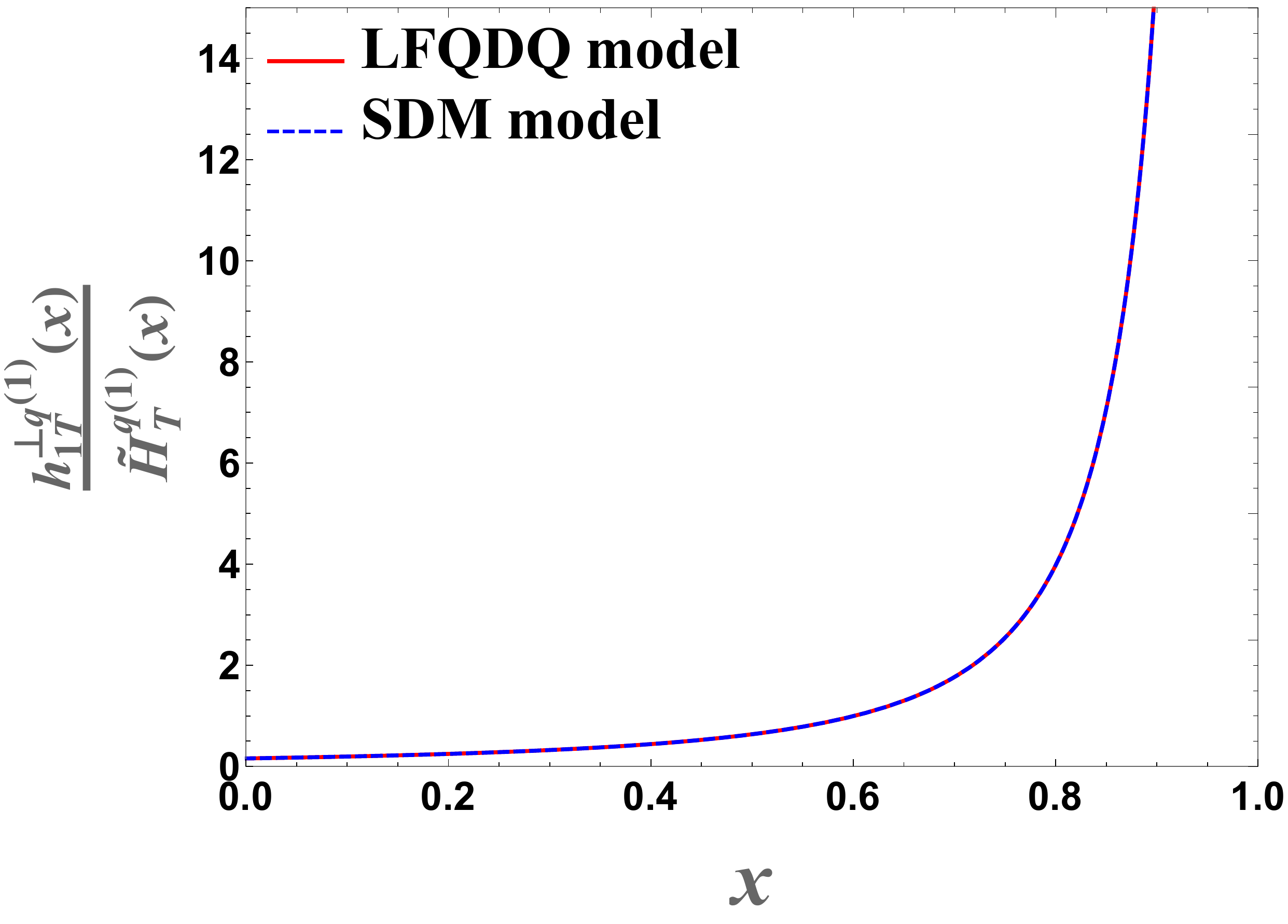}
\caption{ The left plot is corresponding the ratio of the zeroth moment of pretzelosity TMD ($h_{1T}^{\perp u(0)}(x)$) to the corresponding GPD, $\tilde{H}_{T}^{u}(x,0,0)$ . The right one is the ratio of the $h_{1 T}^{\perp u(1/2)}(x)$ to $\tilde{H}_{T}^{u (1/2)}(x)$, and  The lower one is the ratio of the $h_{1 T}^{\perp u(1)}(x)$ to $\tilde{H}_{T}^{u (1/2)}(x)$}.
\label{figpret}
	\end{figure}

	The general structure of the relations in Eqs. (\ref{sdmstmd0}-\ref{sdmstmd1}) and the relations in Eqs. (\ref{pret0}-\ref{pret1}) is not same due to 
	FSI contribution to the T-odd TMDs, like the Sivers and Boer-Mulders function. The pretzelosity distribution is a T-even quantity and at the level of one gluon exchange does not receive contribution from the Wilson line. For this reason,   coupling constants  appear in the relations involving  T-odd TMDs.  Also the relative  power of $(1-x)$ between the moments of the TMDs and of GPDs differs in these two types of relations. In Fig. \ref{figpret} we have compared LFQDQ results in Eqs. (\ref{pret0}-\ref{pret1}) to the scalar diquark model results  in Eq.(\ref{sdmpret0}-\ref{sdmpret1}).
	
The  $n$-th moment of the GPD  $\tilde{{H}}_{T}^{q}$ can be written in terms of  the second derivative of the impact parameter distribution $\tilde{\mathcal{H}}_{T}^{q}$ \cite{Meissner:2007rx}.  Here we consider $n=1$ to have 
		\begin{eqnarray}
			\int d^{2} \mathbf{b}_{T} \frac{\mathbf{b}_{T}^{2}}{2 M^{2}} 2\left(\tilde{\mathcal{H}}_{T}^{q}\left(x, \mathbf{b}_{T}^{2}\right)\right)^{\prime \prime}
			&=&-\pi \int_{0}^{\infty} d b_{T}^{2} \frac{1}{2 M^{2}} 2\left(\tilde{\mathcal{H}}_{T}^{q}\left(x, \mathbf{b}_{T}^{2}\right)\right)^{\prime} \nonumber \\
			&=&\frac{\pi}{M^{2}} \tilde{\mathcal{H}}_{T}^{q}(x, 0) \nonumber\\
			&=& \frac{1}{(2 \pi)(1-x)^{2}} \tilde{H}_{T}^{q(1)}(x)
	\end{eqnarray}
	In the above equation we take the Fourier transform of GPD $\tilde{H}_{T}^{q}(x, t)$ and perform the integration by parts. Then we arrive at the relation
	\begin{eqnarray}
			h_{1 T}^{\perp q(1)}(x)&=&\int d^{2} \mathbf{p}_{T} \frac{\mathbf{p}_{T}^{2}}{2 M^{2}} h_{1 T}^{\perp q}\left(x, \mathbf{p}_{T}^{2}\right) \nonumber\\
			&=&\int d^{2} \mathbf{b}_{T} \frac{\mathbf{b}_{T}^{2}}{2 M^{2}} 2\left(\tilde{\mathcal{H}}_{T}^{q}\left(x, \mathbf{b}_{T}^{2}\right)\right)^{\prime \prime}.
	\end{eqnarray}
	This relation is also valid for both scalar diquark model and quark target model  \cite{Meissner:2007rx}.	
\section{Lensing Function}\label{lensing}
	In \cite{burkardt2002impact}, a nontrivial model dependent relation was found between GPD $E^{q}$ in impact parameter space and the Sivers function $f_{1T}^{\perp q}$. 
	The average transverse momentum of an unpolarized quark in a transversely polarized target \cite{burkardt2007hadron} is defined by
	\begin{eqnarray} 
\left\langle p_T^{q, i}(x)\right\rangle_{U T}
		&=&\int d^{2}\mathbf{p}_T p_T^{i}\Phi^{q}(x,\mathbf{p}_T;S) \nonumber \\
		&=&-\int d^{2} \mathbf{p}_T p_T^{i} \frac{\epsilon_{T}^{j k} p_T^{j} S_{T}^{k}}{M} f_{1 T}^{\perp q}\left(x, \mathbf{p}_T^{2}\right)  \label{sf}\\
		&\simeq & \int d^{2} \mathbf{b}_{T} \mathcal{I}^{q, i}\left(x, \mathbf{b}_{T}\right) \frac{\epsilon_{T}^{j k} b_{T}^{j} S_{T}^{k}}{M}\left(\mathcal{E}^{q}\left(x, \mathbf{b}_{T}^{2}\right)\right)^{\prime}.\label{avgkUT}
	\end{eqnarray}
The above relation (\ref{avgkUT}) is applicable in models where the nucleon state is approximated as an effective two-particle bound state like a diquark model and at the level of one gluon exchange  \cite{burkardt2004sivers}; however it is found to be not valid when vector and axial vector diquarks are included \cite{Pasquini:2019evu}.  Here $\mathcal{I}^{q,i}$ contains the effect of the one gluon exchange in the final state interaction between the active quark and the spectator system.  In such models, Eq. (\ref{avgkUT}) provides the intuitive understanding of the origin of the Sivers transverse SSA. 

\begin{figure}
		\centering
		\includegraphics[scale=0.4,clip]{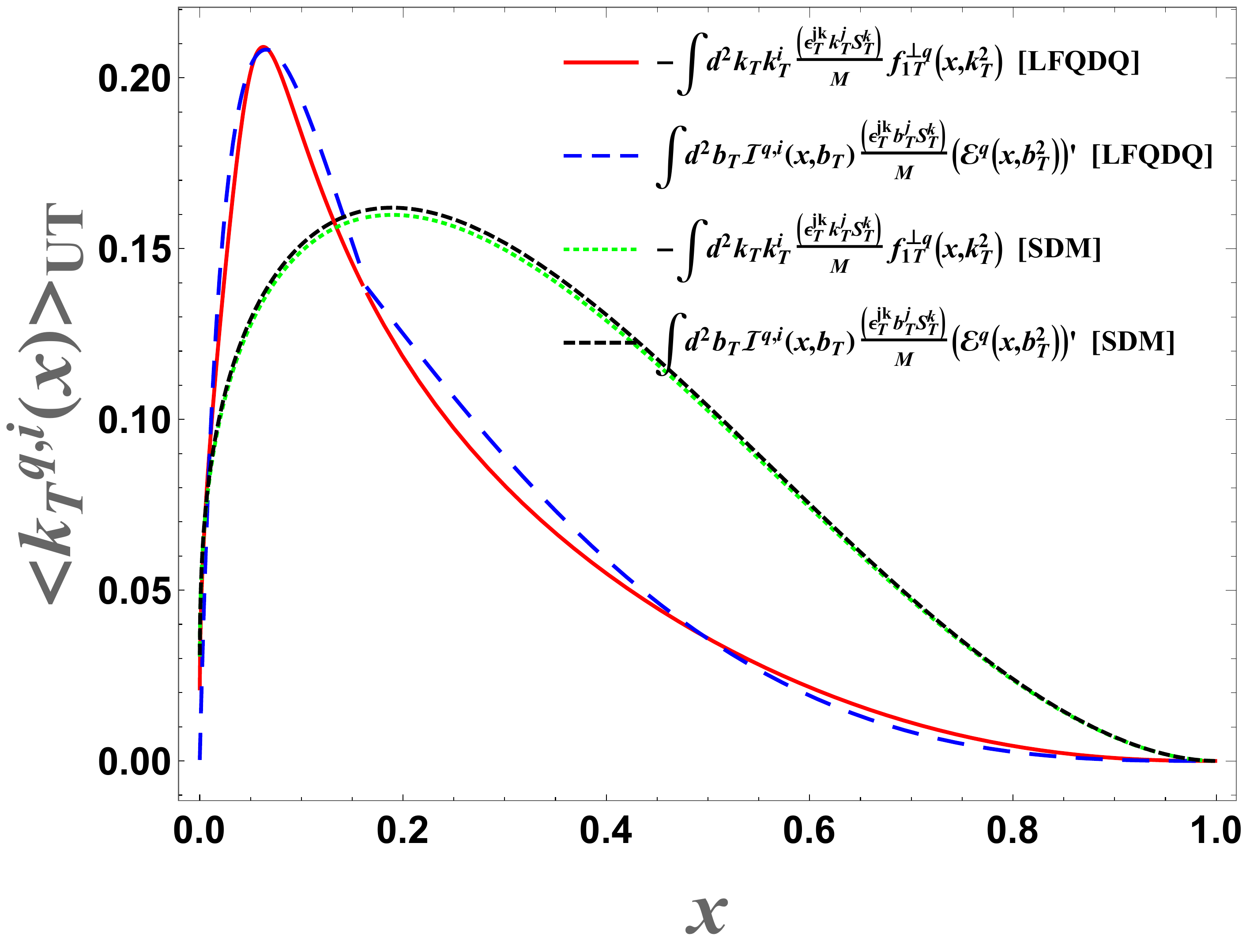}
		\caption{ Lensing function obtained from Sivers function. The average transverse momentum evaluated using Sivers function[Eq.\ref{sf}] and Lensing function [Eq. \ref{avgkUT}] are shown for LFDQDQ and SDM models. For LFQDQ model, the lensing function is given in Eq.\ref{lensing1} and \ref{lensing2}, while the lensing function for the SDM model is given in Eq.\ref{lensingSDM}.}
		\label{figkUT}
	\end{figure}

 However, in general, the average transverse momentum $\left\langle p_T^{q, i}(x)\right\rangle_{U T}$ caused by the Sivers effect can not be factorized into the lensing function $\mathcal{I}^{q,i}$ and the distortion of the impact parameter distribution of quarks in a transversely polarized target which is determined by $(\mathcal{E}^{q})^{\prime}$. So, the relation (\ref{avgkUT}) is model-dependent and  no model-independent relation has been established between the Sivers function $f_{1 T}^{\perp q}$ and GPD $E^{q}$. In our model we have scalar as well as vector and axial vector diquark contributions.  One can still obtain a relation  connecting the Sivers function to a distortion of the GPD $E^q$ in impact parameter space by using an ansatz for the lensing function and obtaining a fit.  Two  different analytic forms of the function are  needed for the low $x$ and high $x$ regions. 
In lower $x$ region $0<x<0.2$ we got the  expression for the lensing function as 
	\begin{eqnarray} \label{lensing1}
		\mathcal{I}^{q,i}(x,\mathbf{b_{T}})=\frac{5 C_F \alpha _S }{2 \pi } x^{3/2} (1-x)^{-1/5} \log^{5}\left(\frac{1}{x}\right) \frac{b_{T}^{i}}{\mathbf{b_T}^{2}} ,
	\end{eqnarray}
	and for the higher $x$ region  i.e. in the region $0.2<x<1$, the lensing function takes the form
	\begin{eqnarray} \label{lensing2}
		\mathcal{I}^{q,i}(x,\mathbf{b_{T}})=\frac{2 C_F \alpha _S }{ \pi } \sqrt{(1-x)} \log(\frac{1}{x}) \frac{b_{T}^{i}}{\mathbf{b_T}^{2}} .
	\end{eqnarray}
	 The lensing function is model dependent but does not depend on the parton type. 
	The corresponding plots for the Eq.(\ref{avgkUT}) by using the Eqs.(\ref{lensing1},\ref{lensing2}) are shown in Fig. \ref{figkUT}.  
	The lensing function in scalar diquark model is relatively easier to extract and  is valid  for  the whole range of $x$( $0<x<1$):
	\begin{eqnarray}
	\mathcal{I}_{SDM}^{q,i}(x,\mathbf{b_{T}})=\frac{5}{\pi} (C_F \alpha_S) (1 - x)  \frac{b_T^i}{b_T^2}.\label{lensingSDM}
	\end{eqnarray}
	In Fig.\ref{figkUT}, the results for scalar diquark model are also shown. The momentum distributions in the two models are not the same, in the scalar diquark model, the distribution peaks at around $x\sim 0.2$ while  inclusion of the axial diquark enhances the diquark contribution and  shifts the peak towards lower $x$.
	
\begin{figure}[htb]
		\centering
		\includegraphics[scale=0.4,clip]{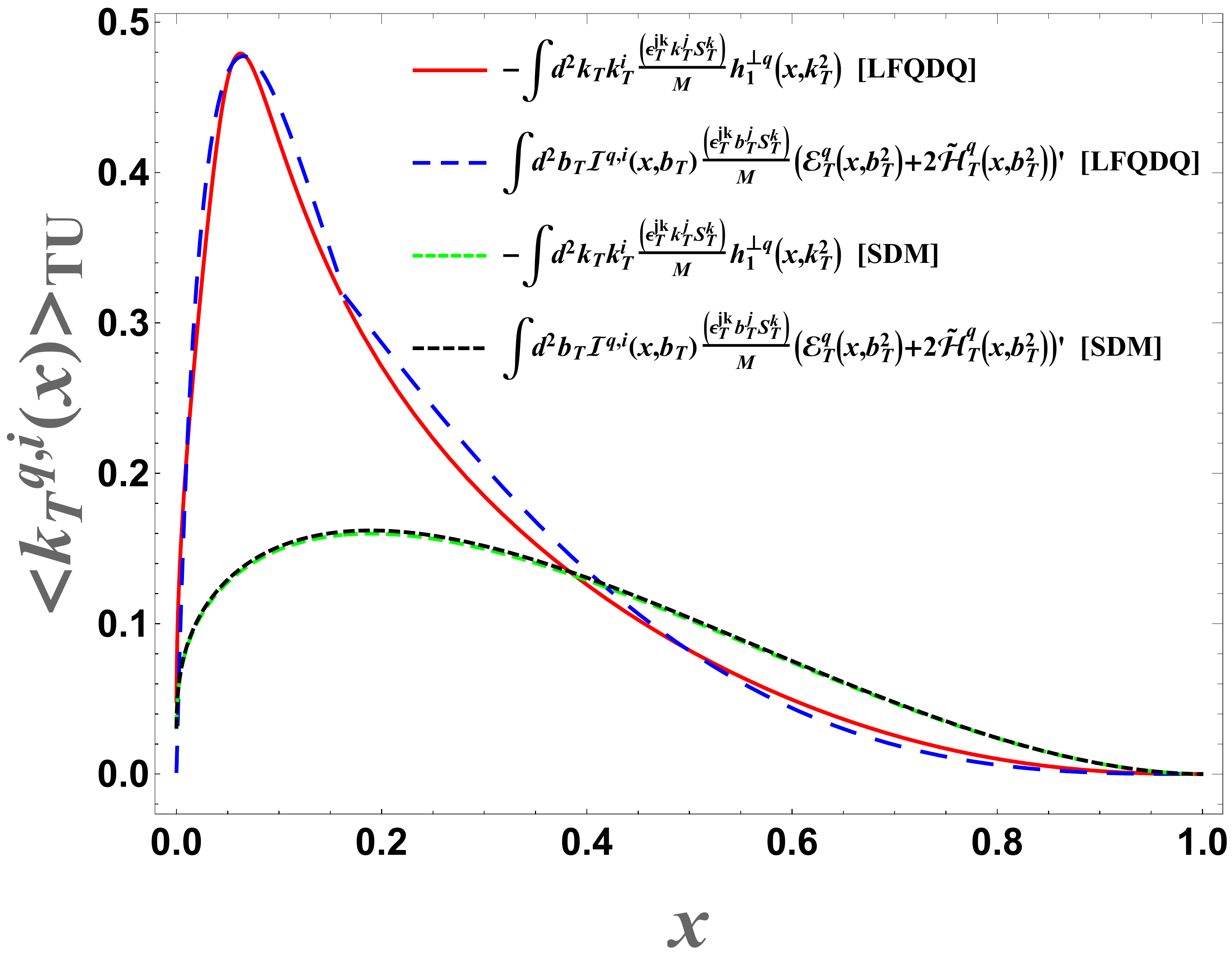}
		\caption{Lensing function for  Boer-Mulders function. The average transverse momentum evaluated using Boer-Mulders function and Lensing function are shown for LFDQDQ and SDM models.  For LFQDQ model, the lensing function is given in Eq.\ref{lensing1} and \ref{lensing2}, while the lensing function for the SDM model is given in Eq.\ref{lensingSDM}}
		\label{figkTU}.
	\end{figure}

The	average transverse momentum $\left\langle p_{T}^{q, i}(x)\right\rangle_{T U}^{j}$ of a transversely polarized
	quark (with polarization along j-direction) in an unpolarized nucleon  can be expressed in terms of the  Boer-Mulders function which is again related with the GPDs\cite{burkardt2004quark}\cite{burkardt2004sivers}.
	\begin{eqnarray}
		\left\langle p_T^{q, i}(x)\right\rangle_{T U}^{j}
		&=&\int d^{2}\mathbf{p}_{T} p_{T}^{i}\Phi^{q,j}(x,\mathbf{p}_T;S) \nonumber \\
		&=&-\int d^{2} \mathbf{p}_T p_T^{i} \frac{\epsilon_{T}^{k j} p_T^{k}}{M} h_{1}^{\perp q}\left(x, \mathbf{p}_T^{2}\right)  \label{bm} \\
		&=& \int d^{2} \mathbf{b}_{T} \mathcal{I}_{\mathrm{SDM}}^{q, i}\left(x, \mathbf{b}_{T}\right) \frac{\epsilon_{T}^{k j} b_{T}^{k}}{M}
		\left(\mathcal{E}_{T}^{q}\left(x, \mathbf{b}_{T}^{2}\right)+2 \tilde{\mathcal{H}}_{T}^{q}\left(x, \mathbf{b}_{T}^{2}\right)\right)^{\prime}. \label{avgkTU}
	\end{eqnarray}
	The Sivers and Boer Mulder's TMDs are the same in this model except the normalization constant and corresponding GPD $E_{T}^{q}+2\tilde{H}_{T}^{q}$ also has the same structure as GPD $E^{q}$.
	The lensing function is also exactly the same as quarks Sivers effect, which means that it is not only independent of the parton type but also of its polarization. The magnitude of the Boer-Mulder function is directly proportional to the distortion of the impact parameter distribution of the transversely polarized quarks inside an unpolarized  nucleon \cite{lu2007connection}. The distortion can be given by the first derivative of $\mathcal{E}_{T}^{q}\left(x, \mathbf{b}_{T}^{2}\right)+2 \tilde{\mathcal{H}}_{T}^{q}\left(x, \mathbf{b}_{T}^{2}\right)$.
	The corresponding plots for the Eqs. (\ref{bm} ) and (\ref{avgkTU}) by using the Eqs.(\ref{lensing1},\ref{lensing2}) for LFQDQ model and Eq.\ref{lensingSDM} for SDM model  are shown  in Figure \ref{figkTU}.

	\section{ Relations among TMDs}\label{relations2}

	The TMDs are found to satisfy many interesting model dependent relations.
	Many such relations have been found for scalar diquark model \cite{Maji:2015vsa}. Though scalar diquark model satisfies saturation limit of Soffer bound, with inclusion of axial-vector diquark, inequality relation  of the Soffer bound is satisfied:
	\begin{eqnarray} \label{morer1}
	h_{1}^{q(s)}\left(x, p_T^{2}\right)<\frac{1}{2}\left(f_{1}^{q(s)}\left(x, p_T^{2}\right)+g_{1}^{q(s)}\left(x, p_T^{2}\right)\right).
	\end{eqnarray}
The ratio of  $g_{1 T}^{q(s)}\left(x, p_T^{2}\right)$ and  $h_{1}^{ q(s)}\left(x, p_T^{2}\right)$ is found to be independent of transverse momentum $p_T$:
	\begin{eqnarray}\label{morer3}
		\frac{ g_{1 T}^{q(s)}\left(x, p_T^{2}\right) } { h_{1}^{ q(s)}\left(x, p_T^{2}\right) }  =\mathcal{J} (x)=
		2 x^{a_{2}^{\nu}-a_{1}^{\nu}-1}(1-x)^{b_{1}^{\nu}-b_{2}^{\nu}}.
	\end{eqnarray}
Two nonlinear relations which connect T-even chirally odd leading twist TMDs in scalar diquark model \cite{Maji:2015vsa} are
	\begin{eqnarray}
		&&\left(g_{1 T}^{q(s)}\left(x, p_T^{2}\right)\right)^{2}+2h_{1}^{ q(s)}\left(x, p_T^{2}\right) h_{1 T}^{\perp q(s)}\left(x, p_T^{2}\right) = 0 , \label{morer7}\\
		&&h_{1}^{q(s)}\left(x, p_T^{2}\right) h_{1 T}^{ \perp q(s)}\left(x, p_T^{2}\right)=-\frac{1}{2}\left[h_{1 L}^{q(s) \perp}\left(x, p_T^{2}\right)\right]^{2}.\label{morer8}
	\end{eqnarray}
	Eq.(\ref{morer8}) implies  that $h_{1}^{q(s)}$ and $h_{1 T}^{q(s)}$ must have opposite signs.  
 The relations with axial-vector diquarks are more involved compared to the relations in scalar diquark model.  The corresponding relations with axial vector diquark  are
	\begin{eqnarray}
		&&\left(g_{1 T}^{q(A)}\left(x, p_T^{2}\right)\right)^{2} + 2h_{1}^{ q(A)}\left(x, p_T^{2}\right) h_{1 T}^{\perp q(A)}\left(x, p_T^{2}\right) =0 \label{morer9}\\
		&&\left[ h_{1 L}^{\perp q(A)} \left(x, p_T^{2}\right) \right]^2 =  -\lambda_1^q  ~h_{1}^{q(A)}(x,p_T^2)h_{1 T}^{\perp q(A)}(x,p_T^{2}), \label{morer12}
	\end{eqnarray}
where the proportionality constant $\lambda_1^q =	\frac{(N_{0}^{\nu 2}-2N_{1}^{q 2})^{2}}{N_{0}^{q 4}} >0$ and the similar conclusion about the opposite polarity of $h_{1}^{q(s)}$ and $h_{1 T}^{q(s)}$ holds. Two more interesting relations with axial vector diquark are
		\begin{eqnarray}
		&&g_{1 T}^{q(A)}(x,p_T^2)=  \frac{N_{0}^{q 2}}{N_{0}^{q 2}+2N_{1}^{q 2}}  h_{1 L}^{\perp q(A)}(x,p_T^2)~~ \Rightarrow ~g_{1 T}^{q(A)}(x,p_T^2)<h_{1 L}^{\perp q(A)}(x,p_T^2),  \label{morer10} \\
		&&\frac{p_T^{2}}{2 M^{2}}h_{1 T}^{\perp q(A)}\left(x, p_T^{2}\right)  =  h_{1 }^{q(A)}\left(x, p_T^{2}\right)-h_{1 T}^{q(A)}\left(x, p_T^{2}\right). \label{morer11}
	\end{eqnarray}
The relations among the TMDs for the full LFQDQ model ( scalar diquark + axial-vector diquark model) can be given as 
\begin{eqnarray}
&&g_{1 T}^{q}(x,p_T^{2}) = \lambda_1^q h_{1 L}^{\perp q}(x,p_T^{2})\\
&&g_{1 T}^{q}(x,p_T^{2}) =\Big[ 2 x^{(-a_{1}^{q}+a_{2}^{q}-1)}(1-x)^{(-b_{1}^{q}+ b_{2}^{q})}\Big]  h_{1}^{q}(x,p_T^{2})\\
&&h_{1T}^{q}(x,p_T^{2}) = \lambda_2^q  f_{1}^{q}(x,p_T^{2})\\
&&h_{1T}^{q}(x,p_T^{2})  + \frac{p_T^{2}}{2M^{2}} h_{1T}^{\perp q}(x,p_T^{2}) = h_{1}^{q}(x,p_T^{2})\\
&&\left( g_{1T}^{q}(x,p_T^{2})\right)^2 + 2 h_{1}^{q}(x,p_T^{2}) h_{1T}^{\perp q}(x,p_T^{2})=0\\
&&\left( h_{1L}^{\perp q}(x,p_T^{2})\right)^{2} =  -2  h_{1}^{q}(x,p_T^{2})h_{1T}^{\perp q}(x,p_T^{2})
\end{eqnarray}
where $\lambda_1^q=\frac{C_{v}^2 N_{0}^{q 2}-3 C_{s}^2 N_{s}^2}{3 C_{s}^2 N_{s}^2+ C_{v}^2 \left(N_{0}^{q 2}-2 N_{1}^{q 2}\right)} ,$  substituting the parameter values we have $\lambda_1^u=0.48, ~\lambda_2^d=1.08$,  and  $\lambda_2^q = \frac{3 C_{s}^2 N_{s}^2-C_{v}^2 N_{0}^{q 2}}{3 C_{s}^2 N_{s}^2+ C_{v}^2 \left(N_{0}^{q 2}+2 N_{1}^{q 2}\right)},$  the values of the constant $\lambda_2^u=0.44, ~\lambda_2^d= -0.93$ imply that for $d$-quark $h_{1T}^{d}(x,p_T^{2})$ is of opposite sign of $ f_{1}^{d}(x,p_T^{2})$.
Similarly the  simple relations among the T-odd (Sivers and Boer-Mulders) TMDs in scalar diquark model
	\begin{eqnarray}\label{morer13}
		h_{1}^{\perp q(s)}\left(x,p_T^{2}\right)=f_{1 T}^{\perp q(s)}\left(x, p_T^{2}\right)
	\end{eqnarray}
	translates into the relation in the scalar and axial vector diquark model as[Eq.(\ref{BM-S})]
	\begin{eqnarray}\label{morer14}
	h_{1}^{\perp q}\left(x, p_T^{2}\right)= \lambda^q f_{1 T}^{\perp q}\left(x, p_T^{2}\right)
	\end{eqnarray}
	where $\lambda^q=\frac{\Big{(}C^2_{S} N_{S}^{q 2}+(\frac{1}{3} N_{0}^{{q}2}+\frac{2}{3}N_{1}^{{q}2})C_{A}^2\Big{)}}{\Big{(}C^2_{S} N_{S}^{q 2}-\frac{1}{3}C_{A}^2 N_{0}^{{q}2}\Big{)}}, \lambda^u=2.29, ~\lambda^d=-1.08$.
	Note that for only scalar diquark,  the Sivers and the Boer-Mulders functions  are same which changes when axial vector diquarks are included.
	
	\section{ Quark orbital angular momentum in a proton}\label{oam}
	The orbital angular momentum (OAM) of a quark inside the nucleon plays an important role in the spin sum rule of the nucleon \cite{kuhn2009spin}. One can extract the total quark contribution to the nucleon spin from the combination of the GPDs using Ji's sum rules as
	\begin{eqnarray}
		L_{z}^{q}= 	\frac{1}{2}\int dx x \left[ H^{q}(x,0,0)+E^{q}(x,0,0) \right]
	\end{eqnarray}
	By subtracting the half of the axial charge $\Delta q= \int dx \tilde{H}^{q}(x,0,0)$ which has the physical interpretation as the spin contribution of quarks with flavour $q$ to the nucleon spin. We can extract the orbital angular momentum of the quark as \cite{ji1997gauge}
	\begin{eqnarray} \label{kineticOAM}
L_{z}^{q}= 	\frac{1}{2}\int dx  \left[ x \left( H^{q}(x,0,0)+E^{q}(x,0,0) \right) - \tilde{H}^{q}(x,0,0) \right]
\label{kinetic}
	\end{eqnarray}
	Where $H^{q}(x,\xi,t)$ and $E^{q}(x,\xi,t)$ are unpolarized GPDs and $\tilde{H}^{q}(x,\xi,t)$ is the helicity dependent GPD. GPDs $H^{q}$, $\tilde{H}^{q}$ and $E^{q}$ respectively in LFQDQ model can be given as 
	\begin{eqnarray}
H^{\nu}(x, 0, t)&=&\left(C_s^2 N_s^2+ C_A^2 \left(\frac{N_0^{\text{$\nu $2}}}{3}+\frac{2}{3}N_1^{\nu }{}^2\right)\right) \left( x^{2 a_1^{\nu }} (1-x)^{2 b_1^{\nu }+1}+\frac{\kappa ^2}{M^2 \log (x)} x^{2 a_2^{\nu }-2} (1-x)^{2 b_2^{\nu }+3}\right) \nonumber \\
&& \times\left(1-\frac{|t|}{4 \kappa^{2}} \log (1 / x)\right) \exp \left[-\frac{|t|}{4 \kappa^{2}} \log (1 / x)\right]  \label{Hq}\\
\tilde{H}^{q}(x, 0, t)&=&\left(C_s^2 N_s^2+ C_A^2 \left(\frac{N_0^{\text{$\nu $2}}}{3}-\frac{2}{3}N_1^{\nu }{}^2\right)\right) \left( x^{2 a_1^{\nu }} (1-x)^{2 b_1^{\nu }+1}-\frac{\kappa ^2}{M^2 \log (x)} x^{2 a_2^{\nu }-2} (1-x)^{2 b_2^{\nu }+3}\right)   \nonumber\\
&& \times\left(1-\frac{|t|}{4 \kappa^{2}} \log (1 / x)\right) \exp \left[-\frac{|t|}{4 \kappa^{2}} \log (1 / x)\right] \label{HTq}\\ 
E^{q}(x, 0, t)& =& 2 \left(C_s^2 N_s^2- \frac{1}{3} C_A^2 N_0^{\text{$\nu $2}}\right)x^{ a_1^{\nu }+a_{2}^{\nu}-1} (1-x)^{ b_1^{\nu }+b_{2}^{\nu}+2}\exp \left[-\frac{|t|}{4 \kappa^{2}} \log (1 / x)\right] \label{Eq}
	\end{eqnarray}
	
\begin{figure}[htb]
		\centering
		\includegraphics[scale=0.32]{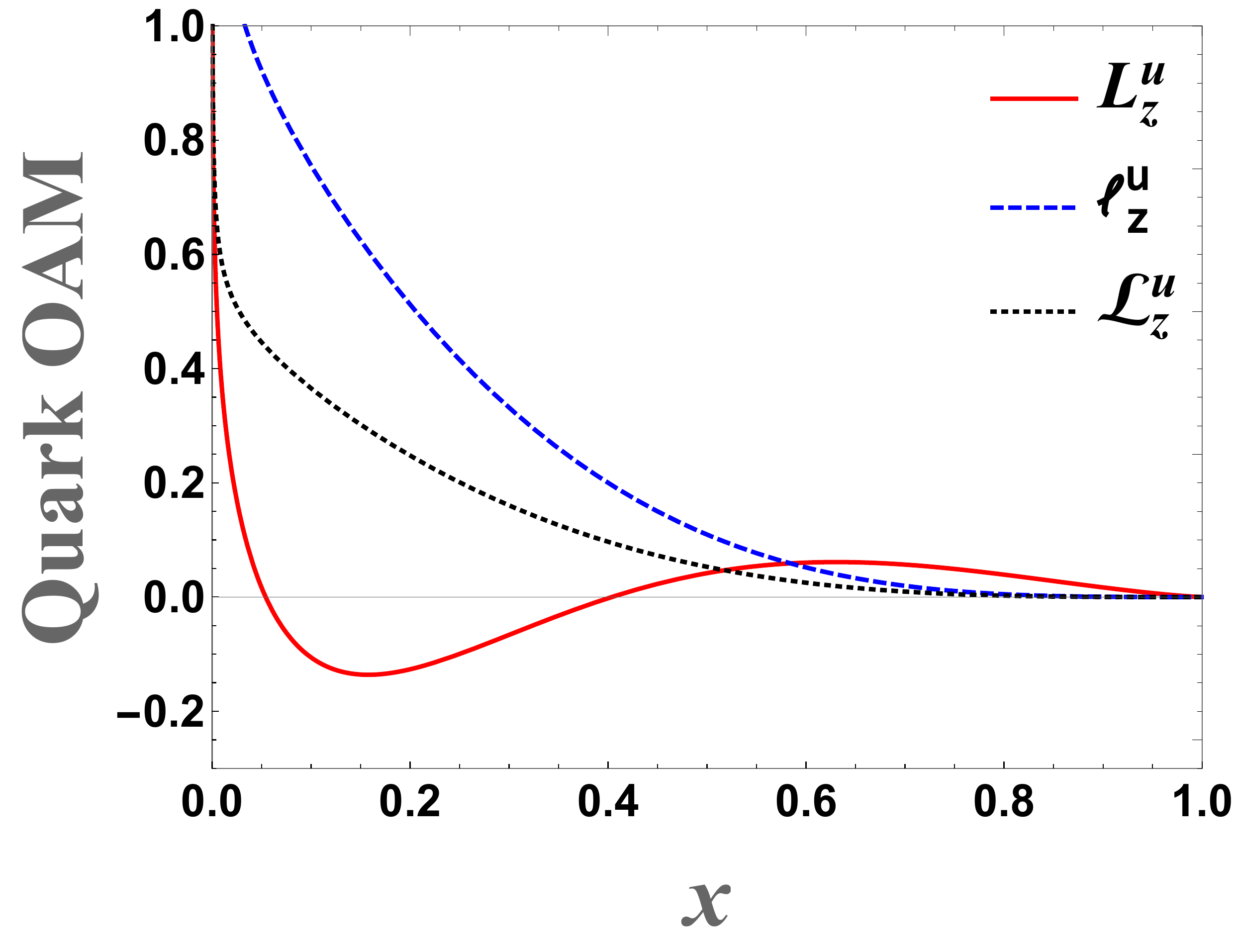}
		\includegraphics[scale=0.32]{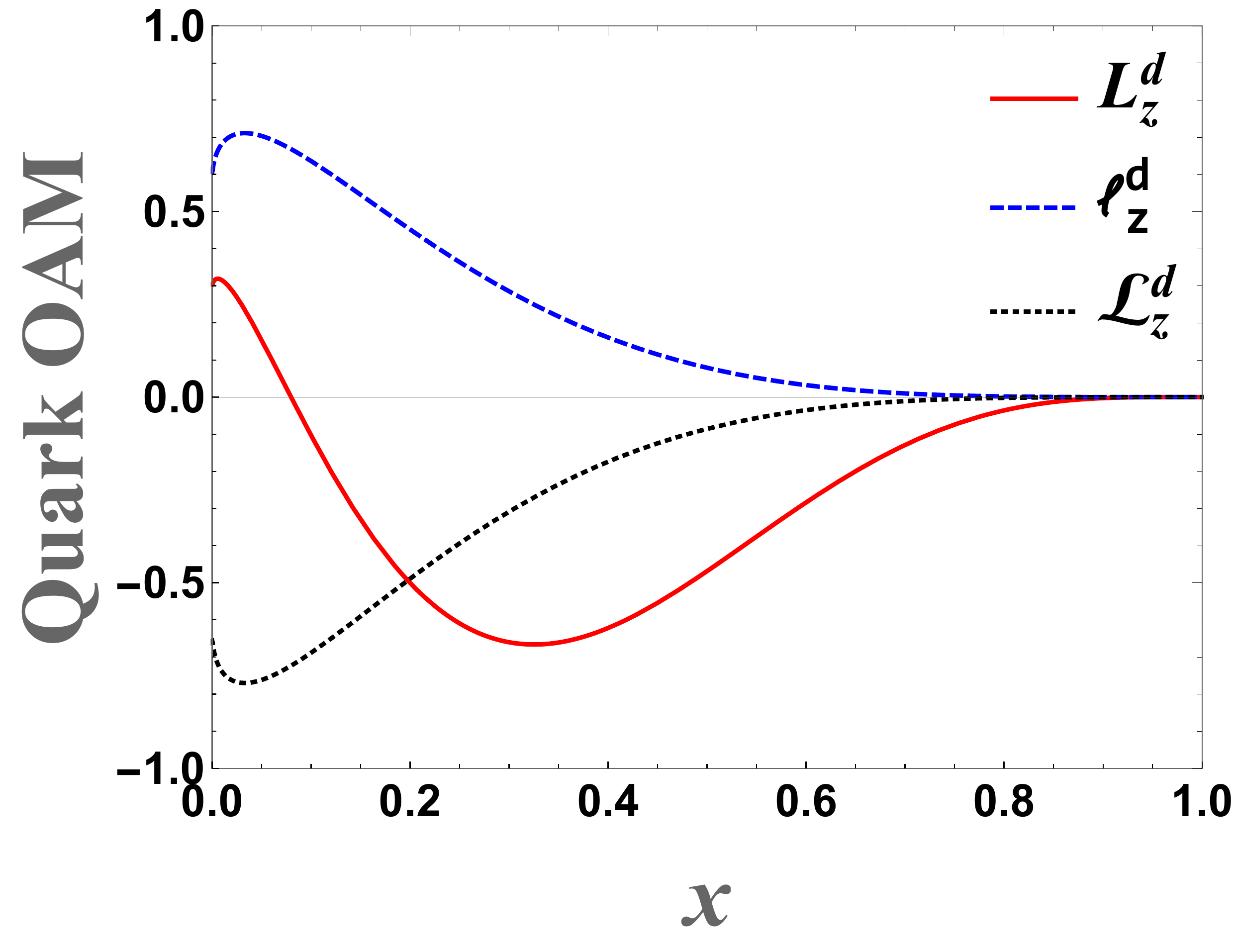}
		\caption{The variation of canonical OAM	$\ell_{z}^{\nu}(x)$ and kinetic OAM $L_{z}^{\nu}(x)$, as well as $\mathcal{L}_{z}^{\nu}$ with longitudinal momentum fraction $x$, for $u$ quark and $d$ quarks.}\label{OAM}
	\end{figure}

The OAM expressed in terms of the GPDs is usually called the kinetic OAM.  
Wigner distributions also contain  the full correlation between quark transverse position and three momentum and one can express the orbital angular momentum in terms of Wigner distribution. This is known as the canonical OAM. 
The average quark OAM in a nucleon polarized in the $z$ direction can be written  as
	\begin{equation}
\hat{\ell}_{z}^{\nu}\left(b^{-}, \mathbf{b}_T, p^{+}, \mathbf{p}_T\right)=\frac{1}{4} \int \frac{d z^{-} d^{2} \mathbf{z}_T}{(2 \pi)^{3}} e^{-i p \cdot z} \bar{\psi}^{\nu}\left(b^{-}, \mathbf{b}_T\right) \gamma^{+}\left(\mathbf{b}_T \times\left(-i \partial_T\right)\right) \psi^{\nu}\left(b^{-}-z^{-}, \mathbf{b}_T\right)
	\end{equation}
	The OAM density operator can be expressed in terms of the Wigner correlator as
	\begin{eqnarray}
\hat{\ell}_{z}^{\nu}=\left(\mathbf{b}_T \times \mathbf{p}_T\right) \hat{W}^{\nu\left[\gamma^{+}\right]}
	\end{eqnarray}
	Thus in the light-front gauge, the average canonical OAM for the quark is written in terms of Wigner distributions as 
	\begin{equation}
	\ell_{z}^{\nu} =\int \frac{d \Delta^{+} d^{2} \Delta_T}{2 P^{+}(2 \pi)^{3}}\left\langle P^{\prime \prime} ; S\left|\hat{\ell}_{z}^{\nu}\right| P^{\prime} ; S\right\rangle
	=\int d x d^{2} \mathbf{p}_T d^{2} \mathbf{b}_T\left(\mathbf{b}_T \times \mathbf{p}_T\right)_{z} \rho^{\nu\left[\gamma^{+}\right]}\left(\mathbf{b}_T, \mathbf{p}_T, x, \hat{S}_{z}\right). \label{oam_wig}
	\end{equation}
	The distribution $\rho^{\nu\left[\gamma^{+}\right]}\left(\mathbf{b}_T, \mathbf{p}_T, x, \hat{S}_{z}\right)$ can be written as\cite{Lorce:2011kd}
	\begin{eqnarray}
\rho^{\nu\left[\gamma^{+}\right]}\left(\mathbf{b}_T, \mathbf{p}_T, x,+\hat{S}_{z}\right)=\rho_{U U}^{\nu}\left(\mathbf{b}_T, \mathbf{p}_T, x\right)+\rho_{L U}^{\nu}\left(\mathbf{b}_T, \mathbf{p}_T, x\right)
	\end{eqnarray}
	where  $\rho_{UU}$ is the  Wigner distribution of unpolarized quark in an unpolarized nucleon and $\rho_{LU}$ is the Wigner distribution of unpolarized  quark in a longitudinally polarized nucleon.
Thus, Eq.(\ref{oam_wig}) can be decomposed into two parts. The term involving $\rho_{UU}$ gives zero:
	\begin{eqnarray}
\int d x d^{2} \mathbf{p}_T d^{2} \mathbf{b}_T\left(\mathbf{b}_T \times \mathbf{p}_T\right)_{z} \rho_{U U}^{\nu}\left(\mathbf{b}_T, \mathbf{p}_T, x\right)=0,
	\end{eqnarray}
	which implies that in an unpolarized nucleon there is no net quark OAM. While the other part can be related to  the twist-2 quark canonical OAM in light-front gauge   and can be written in terms of GTMDs as 
	\begin{eqnarray} \label{canonicalOAM}
\ell_{z}^{\nu}=-\int d x d^{2} \mathbf{p}_T \frac{\mathbf{p}_T^{2}}{M^{2}} F_{1,4}^{\nu}\left(x, 0, \mathbf{p}_T^{2}, 0,0\right).
	\end{eqnarray}
	The correlation between the proton spin and quark canonical OAM can be understood from $\ell_{z}^{\nu}$. If $\ell_{z}^{\nu}>0$, means the quark OAM is parallel to the proton spin and  if $\ell_{z}^{q}<0$, then the quark OAM is antiparallel to the proton spin. 		
	In LFQDQ model the GTMD $F_{1,4}^{\nu}\left(x, 0, \mathbf{p}_T^{2}, 0,0\right)$ can be given as 
	\begin{equation} \label{F11}
	F_{1,4}^{\nu}\left(x, 0, \mathbf{p}_T^{2}, 0,0\right)=-
\left(C_{S}^{2} N_{S}^{2}+C_{V}^{2}\left(\frac{1}{3} N_{0}^{2}-\frac{2}{3} N_{1}^{2}\right)\right)^{\nu}\frac{1}{16\pi^{3}}\frac{(1-x)}{x^2}|A_{2}^{\nu}(x)|^2\exp[-a(x)\mathbf{p}_T^{2}]
	\end{equation} 
	and by using the Eq. (\ref{canonicalOAM}) the canonical OAM can be written as
	\begin{eqnarray}
\ell_{z}^{\nu}(x)=\left(C_{S}^{2} N_{S}^{2}+C_{V}^{2}\left(\frac{1}{3} N_{0}^{2}-\frac{2}{3} N_{1}^{2}\right)\right)^{\nu} \frac{\kappa^{2}}{M^{2}\log(1/x)}x^{2a_{2}^{\nu}-2}(1-x)^{2b_{2}^{\nu}+4}
	\end{eqnarray}
	while the kinetic OAM Eq. (\ref{kineticOAM}) can be calculated from Eqs. (\ref{Hq},\ref{HTq},\ref{Eq}). In this model we get  $L_{z}^{\nu}<0$ for both the up and the down quarks. 
	   The canonical OAM, $\ell_{z}^{\nu}>0$ at $\mu_{0}=0.313$ GeV.  This means that  the quark OAM is parallel to the proton spin for both u and d quarks. Note also that in a scalar diquark model \cite{Chakrabarti:2016yuw} with AdS/QCD wave functions, the OAM is found to be positive for both quarks. This result is model dependent and may be due to the particular form of the AdS/QCD wave functions. The variations of the canonical quark OAM $\ell_{z}^{\nu}(x)$ and kinetic quark OAM, as given in Eq. (\ref{kinetic}) $L_{z}^{q}(x)$ with the longitudinal momentum fraction $x$  are shown in Fig. \ref{OAM} for both $u$ and the $d$ quarks.
	   
A  few interesting points about the OAM in spectator type models are to be noted. In a model without gluons, kinetic and canonical OAM are expected to be equal \cite{Lorce:2011kd}, as the difference between these two are expressed in terms of a gauge potential. The quark OAM was investigated in a simple spectator model with scalar and axial vector diquarks in \cite{Liu:2014zla}. It was found that the relation between the kinetic OAM and the gravitational form factors is not valid in such models, which implied that the kinetic OAM of the quarks is not given by  Eq. (\ref{kinetic}) in such models. In fact, we observed that the kinetic and canonical OAM are not equal in our model, which  
means that a similar conclusion may be drawn here as well. 

In some models,  the pretzelosity $h_{1 T}^{\perp q}(x,p_{T}^{2})$ (\ref{pretzelositytmd}) distribution is also related to the quark OAM  \cite{she2009h,avakian2010transverse}
	\begin{eqnarray} \label{pretOAM}
		\mathcal{L}_{z}^{q}=-\int dx d^{2} \mathbf{p}_T \frac{\mathbf{p}_T^{2}}{2M^{2}} h_{1 T}^{\perp q}(x,\mathbf{p}_{T}^{2}),
	\end{eqnarray}
which in the LFQDQ model has the form 
	\begin{eqnarray} \label{pretlz}
\mathcal{L}_{z}^{\nu}(x)=\left(C_{S}^{2} N_{S}^{2}-\frac{1}{3}C_{A}^{2} N_{0}^{\nu 2}\right) \frac{\kappa^{2}}{M^{2}\log(1/x)}x^{2a_{2}^{\nu}-2}(1-x)^{2b_{2}^{\nu}+4}.
	\end{eqnarray}
	The variation of the $\mathcal{L}_{z}^{\nu}$ with $x$ is shown in Fig \ref{OAM} for both $u$ and $d$ quarks and compared with the OAM through GTMDs Eq. (\ref{canonicalOAM}). In LFQDQ model we got $\mathcal{L}_{z}^{u}>0$ for the up quark and $\mathcal{L}_{z}^{d}<0$ for down quarks \ref{table2}. While for the scalar diquark model the OAM $\mathcal{L}_{z}^{\nu}>0$ for both the up and down quarks respectively. The pretzelosity distribution was found not to be related to the quark OAM in a diquark model with axial vector diquarks \cite{Liu:2014zla} , in our model also, as seen in Fig. \ref{OAM},   we find that is it different from the canonical OAM.  	
	
	Wigner distributions also allow us to study the correlation between spin and OAM of the quark, which is given by the operator \cite{Lorce:2011kd}
	\begin{equation}
C_{z}^{\nu}\left(b^{-}, \mathbf{b}_T, p^{+}, \mathbf{p}_T\right)=\frac{1}{4} \int \frac{d z^{-} d^{2} \mathbf{z}_T}{(2 \pi)^{3}} e^{-i p \cdot z} \bar{\psi}^{\nu}\left(b^{-}, \mathbf{b}_T\right) \gamma^{+} \gamma^{5}\left(\mathbf{b}_T \times\left(-i \partial_T\right)\right) \psi^{\nu}\left(b^{-}-z^{-}, \mathbf{b}_T\right).
	\end{equation}
It can be expressed either in terms of the Wigner distributions $\rho_{UL}^{\nu}$, or equivalently in terms of GTMD as 
\begin{equation}
	C_{z}^{\nu} =\int d x d^{2} \mathbf{p}_T d^{2} \mathbf{b}_T\left(\mathbf{b}_T \times \mathbf{p}_T\right)_{z} \rho_{U L}^{\nu}\left(\mathbf{b}_T, \mathbf{p}_T, x\right) \\
	=\int d x d^{2} \mathbf{p}_T \frac{\mathbf{p}_T^{2}}{M^{2}} G_{1,1}^{\nu}\left(x, 0, \mathbf{p}_T^{2}, 0,0\right).
\end{equation}
The GTMD 	$G_{1,1}^{\nu}\left(x, 0, \mathbf{p}_T^{2}, 0,0\right)$ in LFQDQ model  has the form
\begin{equation} \label{G11}
	G_{1,1}^{\nu}\left(x, 0, \mathbf{p}_T^{2}, 0,0\right)=-
	\left(C_{S}^{2} N_{S}^{2}+C_{V}^{2}\left(\frac{1}{3} N_{0}^{2}+\frac{2}{3} N_{1}^{2}\right)\right)^{\nu}\frac{1}{16\pi^{3}}\frac{(1-x)}{x^2}|A_{2}^{\nu}(x)|^2\exp[-a(x)\mathbf{p}_T^{2}].
\end{equation} 
For $C_{z}^{\nu}>0$ the quark spin and the quark OAM are aligned, while for $C_{z}^{\nu}<0$ they are anti-aligned to each other. From Eq. (\ref{G11}), we calculate $C_{z}^{\nu}$ at $\mu_{0}=0.313$ GeV. In LFQDQ model the numerical values $C_{z}^{u}=-0.284$ and $C_{z}^{d}=-0.234$ for $u$ and $d$ quarks respectively. $C_{z}^{\nu}<0$ implies that the quark OAM is antiparallel to the quark spin, as observed in the scalar diquark model \cite{Chakrabarti:2016yuw}, whereas in the light-cone constituent quark model \cite{Lorce:2011kd}, the $C_{z}^{\nu}$ values are found to be positive for both $u$ and $d$ quarks. The numerical values of the kinetic OAM Eq.(\ref{kineticOAM}), and the canonical quark OAM Eqs.(\ref{pretOAM},\ref{canonicalOAM}) for the up and down quarks in LFQDQ model are given in Table \ref{table2}.
The spin contribution of the quark to the proton spin is defined  as 
\begin{eqnarray} \label{sznu}
	s_{z}^{\nu} =\frac{1}{2} g_{A}^{\nu}=\frac{1}{2} \int d x \tilde{H}^{\nu}(x, 0,0)
	=\frac{1}{2} \int d x d^{2} p_T G_{1,4}^{\nu}\left(x, 0, \mathbf{p}_T^{2}, 0,0\right),
\end{eqnarray}
where $g_{A}^{\nu}$ is the axial charge, and   the GTMD
\begin{eqnarray}
G_{1,4}^{\nu}\left(x, 0, \mathbf{p}_T^{2}, 0,0\right)&=&\left(C_{S}^{2} N_{S}^{2}+C_{V}^{2}\left(\frac{1}{3} N_{0}^{\nu 2}-\frac{2}{3} N_{1}^{\nu 2}\right)\right)\frac{1}{16\pi^{3}}\left[|A_{1}^{\nu}(x)|^2-\frac{\mathbf{p}_T^{2}}{M^{2}x^{2}}|A_{2}^{\nu}(x)|^2\right] \nonumber\\
&&~~~~~\times \exp[-a(x)\mathbf{p}_T^{2}].
\end{eqnarray}
\begin{table}	
	\centering
	\begin{tabular} {|c|c| c|} 
		\hline \hline
		q &   u &   d \\ 
		\hline
		$\ell_{z}^{q}$ Eq.(\ref{canonicalOAM})  & 0.256 & 0.201 \\
		$L_{z}^{q}$ Eq.(\ref{kineticOAM})     & -0.410 & -0.592 \\ 
		$\mathcal{L}_{z}^{q}$ Eq.(\ref{pretOAM})     & 0.124 & -0.218\\ 
		\hline\hline
		\end{tabular}
		\caption{In the light-front AdS/QCD axial-vector diquark model, the values of the canonical OAM $\ell_{z}^{\nu}$; $\mathcal{L}_{z}^{\nu}$, and the kinetic OAM $L_{z}^{\nu}$ for the  $ u$, $d$ quarks.} 
	\label{table2}
\end{table}	
 We should note that the axial charges are highly scale dependent and are measured at high energies, whereas the LFQDQ model has a  low initial scale of $\mu_{0}$=0.313 GeV. So we need to consider the scale evolution of the distributions before comparing with the measured data. For the $d$ quark, the axial charge is known to be negative at larger scales. In Ref.\cite{Chakrabarti:2017teq}
the scale evolution of axial charges are given, where it is shown that the axial charges for the $d$ quarks becomes negative for $\mu^{2}\geq0.15$ $GeV^{2}$. In LFQDQ model we got the axial charges for the up quark and the down quark as  $s_{z}^{u}=1.142$ and $s_{z}^{d}=0.340$ at $\mu_{0}=0.313$ GeV.
while at $\mu^{2}=1$ $GeV^2$ the axial charges for the up quark and the down quarks are given by $s_{z}^{u}=0.73$ and $s_{z}^{d}=-0.54$ respectively \cite{Chakrabarti:2017teq}.

A few observations about other calculations in the literature based on diquark models : in  \cite{Bacchetta:2008af}, a more phenomenological version of a diquark model was used, with the inclusion of both scalar and axial vector diquarks. The LFWFs were parametrized using fit of polarized and unpolarized parton distributions (pdfs) data at the lowest scale.  It was found that the momentum sum rule cannot be satisfied in such models. The spin sum rule has not been investigated in this reference.  The quark OAM has  also been investigated in \cite{Gutsche:2016gcd}  using a LFWF obtained from soft wall ADs/QCD; the spin sum rule has not been explored in this model. In our model, that follows a similar approach, the parameters in the LFWFs, that consists of one quark and a diquark,   are obtained from fits to electromagnetic form factors and pdf data at the initial scale. The spin sum rule constraints the total angular momentum of the diquark.   

\section{Conclusions}\label{concl}
	As GPDs and TMDs encode information about the three dimensional structure of the nucleons and their spin and orbital angular momentum, these distributions  are being investigated in different models.  Both GPDs and TMDs are not physical observables, but many quantities like orbital angular momentum, average momentum of a parton etc can be related to different TMDs and GPDs.  Though there is no one to one correspondence between TMDs and GPDs, they satisfy many interesting relations. Except few, most of the relations are model dependent. In this work we have explored the possible relations in a light front quark-diquark model of the proton. An analytic formula for the lensing function has been formulated in this model.  The lensing function is  model dependent but  is independent of quark flavor and is  the same for unpolarized quarks in a transversely polarized proton or transversely polarized quarks in an unpolarized proton.  Different  types of relations with GPDs and TMDs and their moments are discussed in this model. These relations are  important for model building of the distribution functions for the nucleons. We also calculated the quark orbital angular momentum using different relations. The results are compared with other calculations in similar models. It is observed that the kinetic OAM in this model is not the same as the canonical OAM, which is an indication that the total angular momentum of the quark is not expressed in terms  of the gravitational form factors, as was also observed earlier in the literature in a diquark spectator model. The pretzelosity distribution in this model also does not give the quark OAM.   
	
	{\bf Acknowledgements}
	
	This work is financially supported by  Science and Engineering Research Board  under the Grant No. CRG/2019/000895. 
	
\bibliographystyle{unsrt}
	\bibliography{ref.bib}
\end{document}